\documentclass[amsmath,amssymb,aps,twocolumn, groupedaddress, notitlepage]{revtex4-1}
\usepackage{graphicx}
\usepackage{bm}
\usepackage[normalem]{ulem}
\usepackage{xcolor}

\begin{document}

\preprint{APS/123-QED}
\newcommand{\pto}{ PbTi$\text{O}_3$ }
\newcommand{\sto}{ SrTi$\text{O}_3$ }
\newcommand{\sro}{ SrRu$\text{O}_3$ }
\newcommand{\bto}{ BaTi$\text{O}_3$ }
\newcommand{\red}[1]{\textcolor{red}{#1}}
\newcommand{\blue}[1]{\textcolor{blue}{#1}}
\newcommand{\xx}{ \bm{x} }
\newcommand{\ang}{ \mathrm{\AA} }

\title{Ab Initio  Generalized Langevin Equation }
 
\author{Pinchen Xie}
\affiliation {Program in Applied and Computational Mathematics, Princeton University, Princeton, NJ 08544, USA}

\author{Roberto Car}
\affiliation {Department of Chemistry, Department of Physics, Program in Applied and Computational Mathematics, Princeton Institute for the Science and Technology of Materials, Princeton University, Princeton, NJ 08544, USA}
\author{Weinan E}
\affiliation{AI for Science Institute, Beijing, China,\\
Center for Machine Learning Research and School of Mathematical Sciences, Peking University, Beijing, China}
\date{\today}
\begin{abstract}
     We introduce a machine learning-based approach called ab initio generalized Langevin equation (AIGLE) to model the dynamics of slow collective variables in materials and molecules. In this scheme, the parameters are learned from atomistic simulations based on  ab initio quantum mechanical models. Force field, memory kernel, and noise generator are constructed in the context of the Mori-Zwanzig formalism, under the constraint of the fluctuation-dissipation theorem. Combined with deep potential molecular dynamics and electronic density functional theory, this approach opens the way to multi-scale modeling in a variety of situations. Here, we demonstrate this capability with a study of two mesoscale processes in crystalline lead titanate, namely the field-driven dynamics of a planar ferroelectric domain wall, and the dynamics of an extensive lattice of coarse-grained electric dipoles. In the first case, AIGLE extends the reach of ab initio simulations to a regime of noise-driven motions not accessible to molecular dynamics. 
     In the second case, AIGLE deals with an extensive set of collective variables by adopting a local approximation for the memory kernel and retaining only short-range noise correlations. The scheme is computationally more efficient than molecular dynamics by several orders of magnitude, and mimics the microscopic dynamics at low frequencies where it reproduces accurately the dominant far-infrared absorption frequency.
\end{abstract}
\maketitle

\section{Introduction}
Developing accurate and reliable meso-scale physical models is a long-standing problem ~\cite{muller2002coarse, provatas2011phase, saunders2013coarse}. In this context, the Mori-Zwanzig formalism~\cite{zwanzig2001nonequilibrium} stands out as a general methodology for constructing effective coarse-grained (CG) models for any set of collective variables (CVs) defined in terms of microscopic degrees of freedom, such as the atomic coordinates.
The idea is to project the dynamics of the microscopic variables on the space of the CVs. Finding an approximate surrogate model for the formal projective dynamics requires knowledge of the free energy as a function of the CVs, and brings in two important new effects: memory, because CV dynamics is generally non-Markovian, and noise, associated with the initial condition for the variables eliminated in the projection process.  These effects are difficult to model. As a consequence, one often resorts to simpler approximations for the effective dynamics, such as the Markovian Langevin equation (LE). 

Combined with Landau free energy models~\cite{landau2013statistical,  hohenberg1977theory, chandra2007landau}, LE has been a popular tool for describing meso-scale dynamical processes.  Well-known examples include the Landau-Lifshitz equation for the evolution of the magnetization in materials~\cite{eriksson2017atomistic}, the Allen-Cahn and the Cahn-Hilliard equations for the dynamics of phase transitions and separations ~\cite{allen1979microscopic, cahn1958free}, and, more generally, the phase field models~\cite{chen2002phase} and the phase-field-crystal  models~\cite{PhysRevE.70.051605, PhysRevE.79.035701} for a variety of problems in materials science. Landau-based LEs provide invaluable physical insight but may lack the flexibility required to quantitatively model the CG dynamics of real systems. 
A main issue is the insufficient separation of time scales between the CVs and the noise. In realistic systems, noise may originate from vibrational modes that are not significantly faster than the CVs. In this scenario, the non-Markovian generalized Langevin equation (GLE) is a much better approximation. It can be rigorously derived within the Mori-Zwanzig formalism for Hamiltonians that depend quadratically on the microscopic degrees of freedom~\cite{zwanzig2001nonequilibrium}. In the presence of anharmonicity, the GLE is not exact but can be a flexible enough tool for connecting micro- and meso-scale dynamics, similar in spirit to the way in which semi-local density functional theory (DFT) bridges electronic quantum mechanics and atomistic models~\cite{PhysRevLett.55.2471}. So far, efforts to develop quantitatively accurate GLE models have been limited by difficulties in the parameterization of the memory and noise terms~\cite{wan1995gle, gordon2009gle, satija2019gle}. In the context of bottom-up multi-scale modeling, these difficulties lie in the lack of microscopic data, on the one hand, and of robust algorithms to parameterize the GLE, on the other. 

In recent years, machine learning has emerged as a powerful tool in the study of static and dynamic statistical properties of molecular systems~\cite{PhysRevLett.98.146401,refgap, soap, PhysRevLett.120.143001,schnet,  doi:10.1021/acs.jctc.9b00181, batzner20223}, enabling {\it ab initio} simulations of unprecedented scale \cite{LU2021107624,jia2020pushing}. Today, massive amounts of data can be generated by all-atom molecular dynamics (MD) trajectories with {\it ab initio} accuracy. As we will demonstrate below, machine learning can also address the second difficulty mentioned above.

In this paper, we introduce a machine learning-based method for constructing accurate coarse-grained GLE models from fine-grained/microscopic Hamiltonians. We illustrate the approach with atomistic models derived from DFT, but the methodology can also be applied to microscopic models derived phenomenologically. In our scheme, memory is of finite length and translationally invariant in time, and the noise satisfies the constraint imposed by the second fluctuation-dissipation theorem (2FDT) \cite{kubo1966fluctuation}, which connects the memory kernel to the autocorrelation function (ACF) of the noise. The 2FDT is essential to describe the dynamics of near-equilibrium physical systems.
We call the schemes constructed in this way {\it ab initio} generalized Langevin equation (AIGLE) models, because they are trained on data generated with an 
{\it ab initio} microscopic model. The LE model derived from AIGLE by taking the Markovian limit in the memory kernel and the noise will be called {\it ab initio} Langevin equation (AILE) model.

Previous works have studied data-driven parameterizations of the GLE~\cite{ MCCOY1975431, harris1990I, harris1990II, Berkowitz1981, Berkowitz1983, adelman1983chemical,  Horenko2007, fricks2009time, ceriotti2009langevin, ceriotti2009nuclear, ceriotti2010colored, ceriotti2010efficient, davtyan2015dynamic, lei2016data, Santos2021, russo2022machine}, the LE~\cite{hummer2005position, schaudinnus2015multidimensional, lickert2021data}, the far-from-equilibrium GLE ~\cite{meyer2020non, meyer2021numerical}, and generic stochastic processes~\cite{chorin2015discrete, ma2018model}. 
For GLE restricted by 2FDT, the noise generator is usually constructed from a pre-determined memory kernel or from the ACF of the noise. For instance, in Ref.~\cite{harris1990I, harris1990II}, the noise generator is a Yule-Walker linear autoregressive (AR) model fitted to the ACF of the noise. The resulting model does not guarantee the stationarity of the noise. That can be imposed by adjusting the roots of the characteristic equation, but this may lead to uncontrolled errors. 
Another approach, reported in Ref.~\cite{Berkowitz1981, Berkowitz1983}, assumes that the noise generator is a Fourier series with random coefficients sampled from a distribution defined by the memory kernel. In practice, the Fourier series is of finite length, and the generated noise and its ACF become periodic. 
Recently, Ref.~\cite{lei2016data} proposes to convert a GLE into Markovian equations of motion for fictitious degrees of freedom, by using a finite order Pad\'e approximant for the memory kernel. The dynamics of the fictitious degrees of freedom is constructed according to the memory kernel while retaining the 2FDT constraint on the noise. In general, approaches that use an average property like the ACF to fix the noise are ``mean-field'' approximations, aiming at consistency with data on 2FDT rather than on presumably less relevant features like higher-order correlations or kurtosis. Concerns have been raised that in some of these approaches, the statistical error of the correlation functions may be amplified in an uncontrolled way~\cite{russo2022machine}. 

Alternatively, one can go beyond ``mean-field'' by adopting a regression approach. For example, Ref.~\cite{chorin2015discrete} introduced a non-linear autoregressive model for generic stochastic processes not constrained by the 2FDT. Refs.~\cite{ma2018model, LIU2023105329} used recurrent neural networks for learning generic dynamical systems, as the non-linear nature of general regression tasks may require sophisticated deep neural network models. However, in specialized but important cases such as near-equilibrium systems, knowledge of the free energy surface (FES) and the 2FDT facilitate the task, making it possible to reproduce the time series with relatively shallow and simple neural network regressors. Then, accuracy, stationarity, and efficiency can be achieved simultaneously. In AIGLE, we strive to optimize these three qualities, while ``mean-field'' approaches essentially compromise accuracy. We constrain the memory kernel with the 2FDT and model the noise with a neural network-based generalized autoregressive (GAR) scheme that can deal with insufficient time-scale separation and anharmonic coupling of the modes. These complications are common in real materials but are often overlooked in toy models. To have an efficient noise generator suited for long-time simulation, we keep the neural network as simple as a compact feed-forward neural network. While most data-driven GLEs assume prior knowledge of the FES, in AIGLE not only the noise but also the FES and the couplings to the driving fields can be parameterized. Moreover, while most existing literature uses a one-dimensional GLE, we introduce a multi-dimensional version of AIGLE, based on a local kernel approximation and a consistent GAR model, which can reproduce not only the one-body but also local two-body correlations. The adopted approximation balances efficiency and accuracy, making it possible to deal with infinite-dimensional, homogeneous CVs. To our knowledge, multidimensional GLEs have only been used so far to study the few-body dynamics of low-dimensional CVs~\cite{li2017computing, jung2018generalized, lee2019}. 
 	
In this paper, we expose the details of AIGLE and demonstrate its effectiveness in an {\it ab initio} multi-scale study of ferroelectric lead titanate (\pto). The scheme is not limited to ferroelectric problems and its mathematical structure can be used in reduced models of general crystalline materials. In the present application the order parameters, i.e., the CVs, depend on the local electric dipole moments associated with the crystalline lattice. These local moments act like lattice spins in ferromagnets, and, as the latter, can be coarse-grained to scalar or vector fields in the continuum limit. Unlike lattice spins of fixed magnitude, the local dipoles fluctuate in both direction and magnitude.
The sum of the local dipoles defines the polarization of the system, which is a typical example of a global order parameter in Landau's theory of symmetry breaking. Damped vibrational modes associated with polar phonons are embedded in the dynamics of the local dipoles, inducing oscillating correlations among the dipoles, a behavior that differs significantly from the diffusive dynamics of Brownian particles, whose velocity ACF decays exponentially with time. 
Hence, the difficulties encountered in CG dipole dynamics are similar to those encountered in realistic multiscale models of materials and macromolecules. 

Specifically, we consider two examples of mesoscale dynamics in crystalline materials: the field-driven dynamics of a planar ferroelectric domain wall treated as a virtual particle in a non-Markovian bath, and the dynamics of extensive local order parameters with translationally invariant interactions. In both cases, AIGLE is trained with atomic trajectories generated at room temperature with the Deep Potential (DP) scheme~\cite{PhysRevLett.120.143001}, a deep learning approach that closely reproduces the quantum mechanical potential energy surface at the DFT level of theory. The microscopic lattice dipoles, rigorously defined in the theory of the electric polarization within DFT ~\cite{zhong1995first, marzari2012maximally}, are represented by an equivariant generalization of the DP model~\cite{PhysRevB.102.041121}.
In the first example, we study the glassy dynamics of a planar 180${}^{\circ}$ domain wall induced by a weak electric field~$\bm{E}=E\hat{z}$ parallel to the polarization of one of the domains. We find that the domain wall shifts by a succession of rare events. At low fields, the domain velocity $v_D$ predicted by AIGLE gradually deviates from its AILE counterpart and from the phenomenological scaling law of Merz~\cite{merz1954domain}, according to which $v_D \propto e^{-E_a/E}$, with constant $E_a$.  Merz's law can be derived from the theory of elastic interface motion~\cite{Chauve, Giamarchi2021} that describes the dynamics with an overdamped Langevin equation. Our results suggest that inertia and memory effects captured by AIGLE play a role in the glassy motion of elastic interfaces under weak applied fields. In the second example, we consider the dynamics of a CG lattice of dipoles in the bulk of a compressively-strained \pto crystal. AIGLE reproduces well self and close-neighbor correlations of the dipoles, and captures approximately the ACF of the time derivative of the polarization, whose Fourier transform yields the far-infrared optical spectrum. AILE fails in this task, but still models correctly the relaxation pattern of a domain structure, when this is driven by surface tension and memory is not important. A CG lattice dynamics of extensive CVs like that provided here by AIGLE or AILE, would be useful, in general, in studies of the dynamics of extended crystal defects and of epitaxial growth of materials. 
 
The paper is organized as follows. In Sec.~\ref{sec-aigle} we introduce the AIGLE formalism.  In Sec.~\ref{sec-application}, we use AIGLE for {\it ab initio} multi-scale modeling of \pto. Specifically, we report in Sec.~\ref{sec-domain} a model for the field-driven motion of a planar domain wall in epitaxial \pto. In Sec.~\ref{sec-latt}, we report an extensive model of CG lattice dynamics. Details of one-dimensional AIGLE are in the Material and Methods section. Details of multi-dimensional AIGLE are in the Supporting Information (SI), which also includes the microscopic models for \pto and other technical details.

\section{The AIGLE model}\label{sec-aigle}
 
The starting point is a microscopic model of molecular dynamics. 
Collective variables (CVs), obtained by coarse-graining the microscopic degrees of freedom when constructing the FES, form a column vector $\bm{x}$. The aim is to eliminate the remaining degrees of freedom, and obtain an accurate dynamic model for the CVs, using the GLE ansatz:
\begin{equation}\label{gle}
\small
    M\frac{d^2\bm{x}(t)}{dt^2} = -\nabla  G(\bm{x}) + \bm{F}(t)  + \int_0^t ds MK(s)\frac{d\bm{x}(t-s)}{dt} + \bm{R}(t).
\end{equation}
Here, $M$ is the effective mass matrix, $G(\bm{x})$ is the FES, the vector $\bm{F}$ comprises the external driving forces, $K$ is the memory kernel matrix, and the vector $\bm{R}$ represents the noise. We define $\bm{v} = \frac{d\bm{x}}{dt}$,  $\bm{a}= \frac{d^2\bm{x}}{dt^2}$, ${ \mathcal{F}} = -\nabla G(\bm{x}) + \bm{F}$, and use the subscript $T$ for the transpose of a vector or a matrix. We shall use the brackets  $\langle\cdots\rangle$ to indicate an average over the equilibrium ensemble at $t=0$. We require  $\langle \bm{R} (t) \rangle=0$, and the orthogonality condition $\langle \bm{R}(t) \bm{v}^T(0)\rangle=0$, from which  
the 2FDT can be derived~\cite{kubo1966fluctuation}. The 2FDT prescribes that, at equilibrium, 
$\langle \bm{v}(0)\bm{v}^T(0) \rangle K^T(s) = -  \langle (\bm{M^{-1}R})(0) (\bm{M^{-1}R})^T(s)\rangle$, relating the memory kernel to the ACF of the noise. In addition, although the noise should not be strictly stationary, $\langle \bm{R}(t_0+t) \bm{R}^T(t_0)\rangle$ should be independent of $t_0$ for sufficiently large $t_0$.
 
In AIGLE, Eq.~(\ref{gle}) is learned from the trajectories of $\bm{x}$. The scheme can use, but does not require, a predetermined FES~\cite{izvekov2005multiscale,barducci2008well,barducci2011metadynamics,valsson2014variational,schneider2017stochastic,zhang2018reinforced,zhang2018deepcg,PhysRevX.10.041034,wang2022efficient}, as all the terms in Eq.~(\ref{gle}) can be learned from adequate trajectory data. The resulting GLE satisfies numerically the  2FDT for the equilibrium ensemble, and the model can be extended to near-equilibrium dynamics. Extensions to general nonlinear dynamics beyond the 2FDT would be possible, but this paper is limited to near-equilibrium situations.  
We show that the scheme can be constructed from large-scale MD simulations of realistic materials models.
First, we formulate AIGLE for a one-dimensional CV and then we generalize it to infinite-dimensional lattice CVs. 

To learn from time series data generated by MD, it is convenient to transform the integro-differential equation (\ref{gle}) into discrete form. 
We assume that the memory kernel is homogeneous, i.e., independent of position ~\cite{vroylandt2022position} and time origin~\cite{schilling2022coarse}, and use $\Delta t$ for the time step of the discretized GLE. Setting $t=0$ for the arbitrary starting time, the current time is $t=n\Delta t$, and we use the notation $f_{(n)}$ to indicate a time-dependent function $f(n\Delta t)$. Then, the discretized form of Eq.~(\ref{gle}) for a one-dimensional CV reads
\begin{equation}\label{eom}
\small
    ma_{(n)}  = -\nabla G(x_{(n)}) + F_{(n)} + \sum_{s=0}^{n-1} m K_{(s+\frac{1}{2})} v_{(n-s-\frac{1}{2})} \Delta t + R_{(n)}.
\end{equation}
Eq.~(\ref{eom}) is propagated with the leapfrog algorithm:
\begin{equation}\label{integrator}
\begin{split}
    v_{ (n+\frac{1}{2})} & = v_{(n-\frac{1}{2})} + a_{(n)}\Delta t, \\
    x_{(n+1)} &=x_{(n)} + v_{(n+\frac{1}{2})}\Delta t.
\end{split}
\end{equation}
This setup allows synchronization with MD data when $\Delta t$ equals an integer multiple of $\delta t$, the integration time step of MD.  More accurate multi-step schemes for integrating stochastic dynamics exist~\cite{Tuckerman1991, Leimkuhler2013}, but here we adopt the simple leapfrog scheme because the autoregressive noise model of AIGLE benefits from a simple discretization scheme. Moreover, in consideration of the errors in the free energy calculations, the errors in autoregression, and the lack of a conservation law for stochastic dynamics, the numerical integration error is a minor issue as long as $\Delta t$ is sufficiently small relative to the shortest vibrational period of the CVs.

The free energy $G$ in Eq.~(\ref{eom}) can be parameterized empirically, e.g., using a polynomial ansatz, or, more generally, it can be represented by a neural network ansatz when dealing with high-dimensional CVs. 
The noise term $\{R_{(n)}\}$ in Eq.~(\ref{eom}) is modeled by a generalized autoregressive (GAR) model 
\begin{equation}\label{ag}
    R_{ (n)} = \sum_{k=1}^{m_A} \phi_{(k)}R_{ (n-k)} + \mu_{ (n)} + \sigma_{ (n)}w_{ (n)},
\end{equation}
where $\{\phi_{(k)}\}$ are Yule-Walker linear autoregressive parameters ~\cite{theodoridis2015machine}.  $\mu_{(n)}$ and $\sigma_{(n)}$ are non-linear functions that depend on the history of the noise.
In our approach, $\mu_{(n)}$ and $\sigma_{(n)}$ are the outputs of a deep neural network whose arguments are $R_{(n-1)},\cdots,R_{(n-m_A)}$. The residual noise $w_{(n)}$ represents the uncorrelated part of the noise on the scale of $\Delta t$, and should be close to Gaussian white noise for the scheme to be successful.  The GAR model becomes a standard AR($m_A$) model~\cite{tsay2005analysis} for constant $\mu_{(n)}$ and $\sigma_{(n)}$. When the time dependence of $\mu_{(n)}$ and $\sigma_{(n)}$ cannot be ignored, GAR outperforms AR in reducing $w_{(n)}$ to an almost ideal white noise upon training with MD data, a crucial property for the time correlation functions of CG dynamics to agree with the data.

Eqs.~(\ref{eom}-\ref{ag}) constitute the AIGLE model. The corresponding AILE model is
$ma_{(n)}  = -\nabla G(x_{(n)}) + F_{(n)} +  m \vartheta v_{(n-\frac{1}{2})}   + w_{(n)}$,
where the friction satisfies $\vartheta=\sum_{s=0}^{n-1} K_{(s+\frac{1}{2})} \Delta t$, and the white noise $w_{(n)}$ is fixed by the Markovian 2FDT.
In AIGLE, the parameters defining $G$ and $F$, the memory kernel, the Yule-Walker model, and the neural networks in the GAR model, are learned from MD trajectories kept near thermal equilibrium by a stochastic environment that mimicks a heat bath. This is a common situation in realistic finite-temperature systems. Memory in these open systems should extend over a finite time interval, specified by the integer $m_K$, i.e., $K_{(s+\frac{1}{2})}=0$ for $s>m_K-1$.  $m_K$ and $m_A$ are {\it a priori} parameters of the same order of the relaxation time of the ACF of the velocities of the CVs. At the beginning of the learning protocol, it is convenient to set $m_A=m_K$ to be several times larger than the relaxation time.
Upon fine tuning, the final values of $m_A$ and $m_K$ get close to the relaxation time of the velocity ACF, and we find empirically that a good choice corresponds to $m_A<m_K$. 
The mass $m$ is fixed by the equipartition theorem. 

The recommended learning procedure involves the three actions outlined below. More details are in the Methods section. In the first action the models for $G$ and the memory kernel $\{K_{(s+\frac{1}{2})}\ |\ s\in [0,m_K-1] \}$ in Eq.~(\ref{eom}) are trained on equilibrated MD data. Static and conservative forces $\{F_{(n)}\}$ are absorbed into $G$. 
Two steps are iterated to self-consistency. In the first step, $G$ is optimized while keeping the memory kernel fixed. The loss function is the mean squared deviation from MD of the model prediction for the force on the CV without including noise effects. This procedure is equivalent to minimizing the noise. In the second step, the memory kernel is optimized, while keeping $G$ fixed, by imposing orthogonality of velocity and noise in the least squared sense.
Self-consistency typically requires a few thousand iterations. 
At the completion of the first action, the noise  $\{R_{(n)}\}$ is defined by subtracting the gradient force, $-\nabla G$, and the memory-dependent friction force, predicted by the model, from the true force acting on the CV in the MD data. Then, we turn to the second action, in which the GAR model is optimized using a maximum likelihood loss function, in which the noise values $\{R_{(n)}\}$ constitute the time-series data. This procedure ends when the residual noise is almost white noise, and the GAR model is numerically stationary. At this point, the GLE is fully determined for equilibrium systems and can be used to model mesoscale dynamics under equilibrium conditions. However, when an external driving force $F$ is present, extra training may be necessary. This is done in a third action, which is only executed if needed. In this procedure, the parameters that define $G$ and $F$ are refined with the loss function used in the first action, while keeping the memory kernel and the GAR model fixed. We tested the above procedure and the 1D AIGLE model on a toy system, the infinite harmonic chain, which can be solved analytically within the Mori-Zwanzig formalism. The results of this validation test, reported in the SI, show that AIGLE reproduces with high accuracy the MD data, the analytical Mori-Zwanzig solution, and the 2FDT.  

Finally, we generalize AIGLE for general lattice problems. First, we reinterpret Eq.~(\ref{gle}) for on-lattice CVs. We let $\xx_i=(x_{i1}, \cdots,x_{id_s})^T$ represent a $d_s$-dimensional local order parameter associated to site-$i$ of a $d_l$-dimensional Bravais lattice with periodic boundary conditions. $L$ is the number of sites in the simulation supercell. By concatenating $\{\xx_i\}$ we define a $d_sL$-dimensional CV $\xx=(\xx_1^T, \cdots, \xx_L^T)^T$.
We let $A=\{A_{ij}\}$ be the $L\times L$ adjacency matrix of the lattice. We use $i\sim j$ to indicate a neighboring pair ($A_{ij}$=1). For large $L$, it is not practical to model a dense $d_sL\times d_sL$ memory kernel matrix $K(s)=\{K_{i\alpha}^{j\beta}(s)\}$ ($\alpha,\beta \in [1,d_s]$) and a long-range correlated $\bm{R}(t)$ that preserve exactly the velocity correlation matrix or the 2FDT intrinsic to the data. The simplest approximation is to assume locality and set $K_{i\alpha}^{j\beta}(s)$ to be equal to zero whenever the indices $i$ and $j$ do not correspond to the same site, i.e., when $i\neq j$.
Even with this drastic approximation, the one-body memory kernel will depend not only on the autocorrelation but also on the cross-correlations of the CVs on different sites. Limiting consideration to close-neighbor correlations, we adopt a variational principle for the optimal memory kernel.
We define the orthogonality tensor $\Omega_{i\alpha}^{j\beta}(t)=\langle R_{i\alpha}(t)v_{j\beta}(0)\rangle$, and define the corresponding orthogonality loss, a functional of $K$, by:
\begin{equation}\label{main:lossfunc}
    \mathcal{L}[K](t_K)=\int_0^{t_K} \sum_{i,j,\alpha,\beta} |\Omega_{i\alpha}^{j\beta}(t)|^2 (\delta_{ij}+A_{ij}) dt.
\end{equation}
For a given cutoff of the memory time $t_K=m_K\Delta t$, the optimal one-body memory kernel $K^*$ minimizes Eq.~(\ref{main:lossfunc}), i.e., $K^*=\mathrm{argmin}_{K} \mathcal{L}[K](t_K)$. We call $K^*$ a ``local kernel approximation'' of the exact many-body memory kernel. It enforces a weak form of the 2FDT. The optimality condition for the memory kernel in the case of 1D AIGLE can be regarded as a special case of Eq.~(\ref{main:lossfunc}). We show, in the SI, that the local kernel approximation can be viewed as a special case of a general variational principle for the orthogonality condition.  In practice, we still adopt the discretized form of Eq.~(\ref{gle}) with the leapfrog scheme. We use a multi-dimensional version of Eqs.~(\ref{eom}-\ref{integrator}), but keep the notation $f_{(n)}$ for a time-dependent function $f(n\Delta t)$. We require $K_{(s+\frac{1}{2})}=0$ for $s\geq m_K$. 
Given the force field $\bm{\mathcal{F}}$, after discretizing Eq.~(\ref{main:lossfunc}), we obtain a least-square solution for the optimal memory kernel $K^*_{(s+\frac{1}{2})}$ for $0\leq s < m_K $.  The derivation, given in the SI, is lengthy but straightforward. We also generalize the 1D GAR model to the multi-dimensional case. Note that, although $K^*$ is one-body, the noise $R$ is still spatially correlated as required by the 2FDT. Thus, the noise generator can not be defined locally as commonly done with molecular dynamics thermostats. To deal with this complication, we allow the noise $(R_{(n)})_i$ at site-$i$ and time step $n$ to depend not only on its own history but also on the history of a finite number of neighboring sites-$j$ defined in terms of the adjacency matrix $A$. The sites-$j$ could include nearest neighbor sites ($A_{ij}=1$), next nearest neighbor sites, and so on. In other words, the multi-dimensional GAR model is analogous to a graph neural network~\cite{gilmer2017neural} on the graph of the CVs defined by $A$. 
The details of the multi-dimensional GAR model are given in the SI. The training of multi-dimensional AIGLE follows the same protocol of uni-variant AIGLE.

\section{Applications}\label{sec-application}
\subsection{Domain wall as a virtual particle}\label{sec-domain}
\begin{figure}
    \centering
    \includegraphics[width=\linewidth]{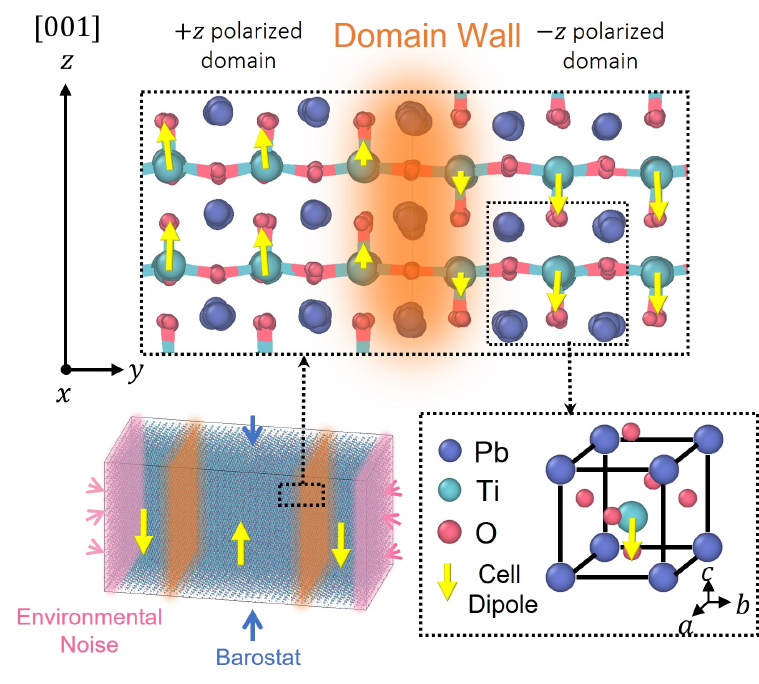}
    \caption{Upper panel: Lateral view of the 180${}^{\circ}$ domain wall in \pto. The bonds between titanium atoms and the nearest oxygen atoms are shown to visualize the domain separation.  The yellow arrows represent the local dipoles, which are weaker near the domain wall. Lower panel left: A $20\times 40\times 20$ supercell of \pto with parallel twin domain walls on the $xz$ plane. The $x$ and $y$ dimensions are fixed to match the experimental lattice constant (see text) while the $z$ dimension fluctuates under constant pressure $P_z=28$kbar (see text) at temperature $T=300$K. Lower panel right: The elementary cell of \pto. }
    \label{fig:atoms}
\end{figure}

In this section, AIGLE is used to study a prototypical problem of ferroelectric domain switching --- the field-driven motion of a twin 180${}^{\circ}$ domain wall in epitaxially grown \pto. Schematic drawings of the domain wall and of the crystal structure are shown in Fig.~\ref{fig:atoms}. Electric dipole moments (local dipoles) $\bm{p_j}$, represented by yellow arrows in Fig.~\ref{fig:atoms},
are associated with the Ti-centered elementary cells-$j$~\cite{meyer2002ab}. The polarization is defined by $\bm{\mathcal{P}}=\sum_j \bm{p_j}/V$, with $V$ the volume of the sample. Polarization changes are experimentally observable. AIGLE is constructed from {\it ab initio} electronic structure data within DFT. Microscopic definitions of $\bm{p_j}$ and $\bm{\mathcal{P}}$ are given in the SI.

\begin{figure*}[tb]
    \centering
    \includegraphics[width=\linewidth]{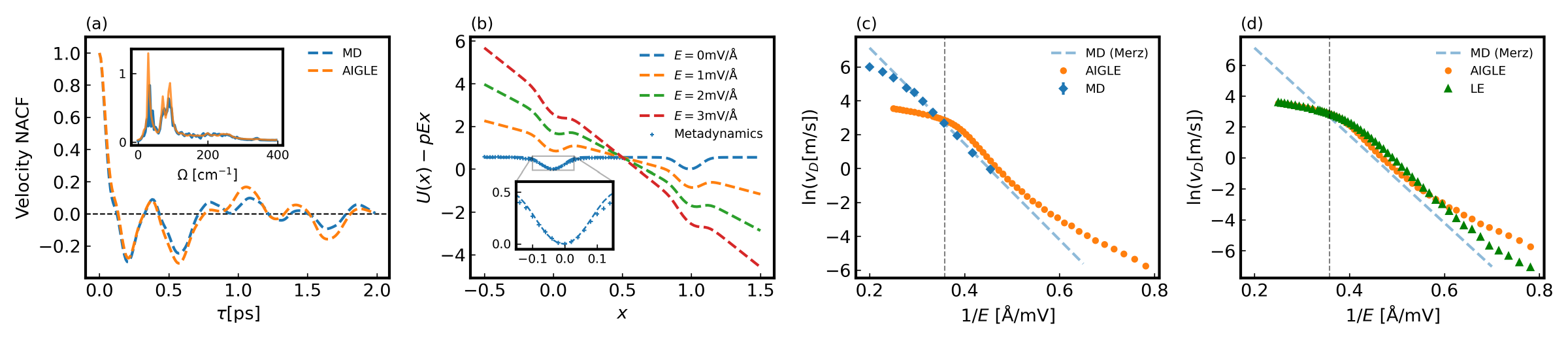}
    \caption{(a) The velocity NACF $C^{\text{MD}}_{vv}(\tau)$ from MD (blue) and the velocity NACF $C^{\text{GLE}}_{vv}(\tau)$ from AIGLE (orange), for $\tau<2$ps. For $\tau>2$ps the correlations are confined to the interval [-0.1,0.1] and decay rapidly to zero. 
    The inset reports $\Re(\hat{C}^{\text{MD}}_{vv}(\Omega))$ (blue) and $\Re(\hat{C}^{\text{GLE}}_{vv}(\Omega))$ (orange) in arbitrary units. The first peak of $\Re(\hat{C}^{\text{GLE}}_{vv}(\Omega))$ is located near $30\mathrm{cm^{-1}}$. (b)
    Free energy profiles along the CV $x$ in the presence of driving fields of various strengths. Two periods of 
    the parametrized $U(x)$ (see text) are shown. $U(x)$ is represented by a dashed blue line. Metadynamics results are also reported as blue crosses, for comparison. The inset shows a magnified plot of the free energy basin including $U(x)$ and metadynamics data.   (c) The natural logarithm of 
    the domain wall velocity $v_D$ is plotted against $1/E$. The vertical dashed line indicates $E=E^*$. Each GLE data point (orange dot) is the average of five AIGLE simulations lasting $0.5\mu$s each. The corresponding error bars are smaller than the size of the dot. Each MD data point (blue diamond) is the average of one hundred MD trajectories lasting $0.1$ns each. The error bars are smaller than the size of the diamond. The dashed blue line is the best fit of Merz's law with MD data ($E_a=28\mathrm{mV/\AA}$). (d) A comparison of AIGLE and LE predictions for $\ln v_D$ vs $1/E$.  Each LE data point (green triangle) is the average of five LE trajectories lasting $0.5\mu$s each. The error bars are smaller than the size of the triangle.
    }
    \label{fig:domain}
\end{figure*}

In epitaxially strained tetragonal \pto, polarized along the $[001]$ crystallographic direction at room temperature ($T=300$K),  the ferroelectric domains have  narrow 180${}^{\circ}$ domain walls  ~\cite{meyer2002ab}. A CV that describes continuously the switch of a domain from $+\hat{z}$ to $-\hat{z}$ is $\alpha = \sum_i \tanh \frac{p_{iz}}{p_*}/2A_{\mathrm{wall}}$, where the sum extends to the local dipoles. We choose a value of $p_*$ that is close to the bulk average of $\|p_{iz}\|$. In the simulations, we set the parameter $A_{\mathrm{wall}}$ to be equal to $400$, the supercell area in the $xz$ plane in units of elementary cells.
With this definition, when the $+\hat{z}$ domain grows by one layer of unit cells in the $y$-direction, the increment of $\alpha$ is approximately equal to $1$. 

We model the motion of the domain wall driven by an external electric field $\bm{E}=E\hat{z}$. Experimentally, the domain wall velocity, $v_D$, obeys approximately a phenomenological law suggested by Merz ~\cite{merz1954domain}, according to which $v_D = v_0 e^{-E_a/E}$, with $v_0$ and $E_a$ empirical parameters. For small $E$, $v_D\rightarrow 0$, and the wall dynamics is glassy. This motion, called domain wall creep, is usually initiated by the nucleation and growth of flat nuclei at the separating interface~\cite{shin2007nucleation}. On a coarser time scale, the moving interface can be viewed as a virtual particle that performs a succession of noise-activated, rare hopping events, rather than a steady continuous motion.  This behavior cannot be deduced from phenomenological laws, like Merz's, and is usually ignored in continuum models, but can be probed, in principle, with microscopic simulations~\cite{liu2016intrinsic}. However, glassy dynamics can easily exceed all-atom simulation capabilities when the time scale is of the order of the microsecond or longer. To cope with the long-timescale bottleneck, one often turns to kinetic Monte Carlo~\cite{shin2007nucleation}, an approach that typically requires {\it ad hoc} iteration rules and assumes Markovianity. AIGLE can simulate non-Markovian dynamics with {\it ab initio} accuracy for time scales comparable to those reachable by kinetic Monte Carlo.  
 
We generate training data for AIGLE
with MD simulations of \pto.  We adopt the Deep Potential (DP) model for the interatomic interactions, and an effective Born charge (BC) model for the local dipoles (see SI). 
The MD supercell is shown in Fig.~\ref{fig:atoms}, where the $x$ and $y$ dimensions are fixed to match the experimental lattice constant $a=3.91\mathrm{\AA}$, and the $z$ dimension is barostatted at a constant pressure of $P_z=28$kbar, a value chosen to roughly match the experimental lattice constant $c$ at 300K and atmospheric pressure (see SI and Ref.\cite{xie2022ab, zhong1995first} for more details). 
With the above setup, we run MD trajectories at $T=300$K, with temperature controlled by a stochastic thermostat, in the presence of homogeneous electric fields of varying magnitude $E$, with $0\leq E \leq 3~\mathrm{mV/ \AA}$, along the $\hat{z}$ direction. The microscopic data for the CV $\alpha$ are extracted from these trajectories. In these simulations, the atomistic degrees of freedom equilibrate quickly with the environment, and the loss of detailed balance is mostly associated with the CV describing domain motion. 
  
The atomistic simulations suggest that the dynamics of $\alpha$ resemble that of a virtual particle subject to colored noise in a tilted periodic potential~\cite{cheng2015long}. When $E$ is small, the particle is trapped in a metastable equilibrium and the velocity ACF of $\alpha$, defined by
$\mathcal{A}_{\dot{\alpha}\dot{\alpha}}(\tau)=\langle \dot{\alpha}(\tau)\dot{\alpha}(0)\rangle$,  exhibits several characteristic oscillations (modes), i.e., a behavior dramatically different from the simple exponential decay characteristic of Brownian dynamics driven by white noise. These characteristic modes originate mainly from the optical phonons of \pto and provide the thermal fluctuations that activate nucleation-driven creep events at small driving fields.  
Taking time scale separation into account, it is convenient, when constructing a CG model, to filter out the frequencies much higher than that of the slowest mode of $\alpha$ ($\approx 30\mathrm{cm^{-1}}$). Then, a new CV, $x$, is constructed
by acting on $\alpha$ with a truncated Gaussian filter in time:
\begin{equation}\label{cgtime}
\small
    x_{(n)} =  \sum_{q=0}^{3\varsigma} \frac{\exp(-\frac{q^2}{2\varsigma^2})}{\sum_{q=0}^{3\varsigma} \exp(-\frac{q^2}{2\varsigma^2})}\alpha(t=n\Delta t - q\delta t).
\end{equation}
Here, $\varsigma$ is the truncation parameter that we set equal to $40$. With this choice, the modes of $\alpha$ in the range $[0,100]~\mathrm{cm^{-1}}$ are barely affected, while the modes with higher frequency are suppressed (see SI). The AIGLE integration time step $\Delta t = 10$~fs is equal to five times the MD time step $\delta t = 2$~fs. This procedure is substantiated by the fact that the residual noise, after training the model, is indeed very close to white noise. The GLE equation of motion for $x$, deriving from Eq.(\ref{eom}), is:
\begin{equation}\label{gle:domain}
\small
        a_{(n)}  = -\partial_x U(x_{(n)}) + pE + \sum_{s=0}^{n-1} K_{(s+\frac{1}{2})} v_{(n-s-\frac{1}{2})} \Delta t + \frac{1}{m}R_{(n)}.
\end{equation}
Here, we parameterized the FES of Eq.(\ref{eom}) with the periodic function $G(x)= mU(x)=mU_b \tanh (k(1-\cos \omega(x-x_0)))$. $m\approx1.0\times10^3$ amu is a scalar mass, estimated with the equipartition theorem.  
$U_b$, the barrier height, is predetermined with metadynamics~\cite{barducci2008well} since it is hard to fit it accurately in the near-equilibrium regime without enhanced sampling. The external force in Eq.~(\ref{eom}) is represented by $F(x)=mpE$. The values of the parameters $k$, $\omega$, $x_0$ and $p$, are fixed by training.  
The time cutoffs for the memory kernel $K$ and for the GAR model for $R$ are $m_K\Delta t=2$ps and  $m_A\Delta t=0.4$ps, respectively. Assuming a linear response regime, the model parameters are independent of $E$. The AIGLE model introduced here is trained on several MD trajectories with $E\in [2.0,2.4]$mV/{\AA}. Details of training and validation can be found in the SI. 
MD systems are at metastable equilibrium for $E \approx 2$mV/{\AA} and near equilibrium for smaller $E$.

Comparison of $C^{\text{MD}}_{vv}(\tau)$, the normalized autocorrelation function (NACF) of the CV velocity extracted from MD, with its AIGLE counterpart, $C^{\text{GLE}}_{vv}(\tau)$, provides a direct validation of AIGLE. The two NACFs, calculated at metastable equilibrium conditions for $E=2$mV/{\AA}, are reported in Fig.~\ref{fig:domain}(a). They agree well with each other but for minor discrepancies. The figure also indicates that the adopted cutoff $m_K$ is large enough to satisfy the condition $C^{\text{MD}}_{vv}(\tau>m_K\Delta t)\ll 1$. A major cause of the small differences between MD and AIGLE is apparent in the inset of Fig.~\ref{fig:domain}(a), which 
reports the real parts of Fourier transforms of the velocity NACFs,  $\Re(\hat{C}^{\text{MD}}_{vv}(\Omega))$ and $\Re(\hat{C}^{\text{GLE}}_{vv}(\Omega))$. The slowest mode occurs at
$\Omega\approx 30\mathrm{cm^{-1}}$ in both NACFs. In $\Re(\hat{C}^{\text{MD}}_{vv}(\Omega))$ this mode exhibits a fast oscillatory line shape, indicating relaxational origin.
The same mode in $\Re(\hat{C}^{\text{GLE}}_{vv}(\Omega))$ has a smooth Lorentzian line shape with a peak frequency that matches the harmonic frequency of the nearly quadratic free energy basin depicted in Fig.~\ref{fig:domain}(b). This indicates that the relaxational fluctuation of the domain wall is turned into an effective harmonic oscillation in a potential well. Since the ansatz for $U(x)$  assumes a smooth, rather than fractal, dependence on $x$, the slowest mode of $\Re(\hat{C}^{\text{GLE}}_{vv}(\Omega))$ displays a clean harmonic peak, sharper than the relaxational peak of $\Re(\hat{C}^{\text{MD}}_{vv}(\Omega))$. This subtle difference is, in fact, a desired consequence of coarse-graining the FES. Two other modes (near $80\mathrm{cm^{-1}}$) are displayed by $\Re(\hat{C}^{\text{GLE}}_{vv}(\Omega))$ and by $\Re(\hat{C}^{\text{MD}}_{vv}(\Omega))$ as well. At higher frequencies the spectrum of $\Re(\hat{C}^{\text{GLE}}_{vv}(\Omega))$ is quite smooth and agrees well with the MD results. 

 
The optimized free energy profile as a function of $x$ is shown in Fig.~\ref{fig:domain}(b).  Metastability disappears for $E$ greater than $E^*\approx2.8$mV/{\AA} when the profile becomes monotonic. 
Hence, $E^*$ represents the threshold beyond which the near-equilibrium regime appropriate for AIGLE is no longer valid. Under near-equilibrium conditions, the lifetime of a metastable state should be much longer than the relaxation times of the atomic vibrations. To study this phenomenology, we run AIGLE for a dense grid of electric field values in the interval $(1,5)$~mV/{\AA}.  
To visualize the variation of the domain velocity $v_D$, which spans several orders of magnitude, we display in Fig.~\ref{fig:domain}(c) the natural logarithm of $v_D$, extracted from AIGLE and MD simulations, respectively, as a function of $1/E$. MD data are only available for $v_D\gtrapprox 1$m/s due to time limits of fully atomistic simulations. 
When AIGLE and MD data are both available, the two approaches agree well for $E \leq E^*$, i.e., under near-equilibrium conditions. For $E > E^*$, 
the domain velocity of MD is significantly larger than its AIGLE counterpart.  From a coarse-graining point of view, this occurs because the 2FDT, valid near equilibrium, has been imposed far from equilibrium. From a microscopic point of view, the electric dipoles, temporarily associated with the moving domain wall, are unable to dissipate energy before separating from the wall. Far from equilibrium memory effects are different from those learned for $E < E^*$. Thus, the present AIGLE model should only be used when $E \leq E^*$, i.e. within the creep regime of the Markovian theory of elastic interface dynamics~\cite{Chauve, Giamarchi2021}. When $v_D\gtrapprox 1$m/s, MD shows linear behavior of $\ln v_D$ with $1/E$, in agreement with Merz's law: $\ln v_D=\ln v_0-E_a/E$~\cite{merz1954domain}. A best fit of the MD data to this law gives $E_a=28\mathrm{mV/\AA}$. 
AIGLE gives essentially the same result, $E_a=27\mathrm{mV/\AA}$, for $E\in [2\mathrm{mV/\AA},E^*]$. Thus, $v_D$ at low electric fields can be estimated from Merz's law fitted to MD for $v_D\gtrapprox 1$m/s, as done, e.g., in Refs.~\cite{shin2007nucleation, liu2016intrinsic}. 
However, when $v_D \ll 1$m/s, direct AIGLE simulations display a gradual deviation from Merz's law, as illustrated in Fig.~\ref{fig:domain}(c). When $1/E>0.6\mathrm{\AA/mV}$,  AIGLE predicts a $ v_D$ higher than Merz's law by orders of magnitude. This behavior is similar to the stretched exponential inferred from the relation $\ln (v_D/v_0)  \propto (E/E_c)^{-\mu}$ of the Markovian theory of elastic interfaces ~\cite{Chauve} when the dynamic exponent $\mu$ is less than $1$. This theory assumes a Markovian overdamped regime. Yet, the deviation from Merz's law, predicted by AIGLE at low fields, is markedly more rapid than the stretched exponential of the Markovian theory.  
This suggests that memory and inertia play an increasingly important role in the regime of very rare domain motions.  

 To gauge the implication of non-Markovianity, we approximate AIGLE with AILE. 
Fig.~\ref{fig:domain}(d), shows that LE predicts for $v_D$ a behavior consistent with Merz's law, which is not surprising because the derivation of Merz's law requires a Markovian approximation. The same figure shows that LE and AIGLE agree well with each other when $1/E$ is close to $1/E^*$, a situation in which the external driving force dominates over memory and noise. For larger $1/E$, the LE predicted behavior deviates from a pure exponential in a very minor way, underestimating $v_D$ by orders of magnitude relative to AIGLE at the largest values of $1/E$. Within Markovian dynamics the friction is always dissipative, hindering thermally-activated motion irrespective of the time scale of the creep events. An even simpler dynamics is postulated in the Markovian theory of elastic interfaces~\cite{Chauve,Giamarchi2021} that adopts an overdamped Langevin equation, where both memory and inertia effects are absent. By contrast, within AIGLE, memory results from a convolution of oscillating functions and can occasionally lead to a kinetic energy increase over a short time interval. In combination with inertia, this effect enhances the likelihood of barrier crossing. From the perspective of transition state theory, this effect can be understood as effectively enhancing the pre-exponential factor in the formula for the rate.  Non-Markovian effects that facilitate barrier-crossing have also been discussed in other contexts, such as, e.g., in the Grote-Hynes theory of chemical reaction rates~\cite{grote1980stable}. 

Using the domain velocity $v_D(E)$ calculated with AIGLE, we can estimate the hysteresis loop observed experimentally when the polarization is reversed by a driving field. We report in the SI a hysteresis loop calculation using a very simple model of ferroelectric switching that ignores point defects and dependence on the curvature of the domain wall. The results are in semi-quantitative agreement with experiments.

\subsection{ Coarse-grained lattice dynamics} \label{sec-latt}
Here, we use multi-dimensional AIGLE to describe the dynamics of lattice CVs, which are either the local dipole moments $\{\bm{p_j}\}$ or a CG model of them. The underlying microscopic model is the all-atom DP model of Sec.~\ref{sec-domain}. For each atomic configuration the local dipoles are provided by a neural network model (see SI). 

 \begin{figure*}[t]
    \centering
    \includegraphics[width=\linewidth]{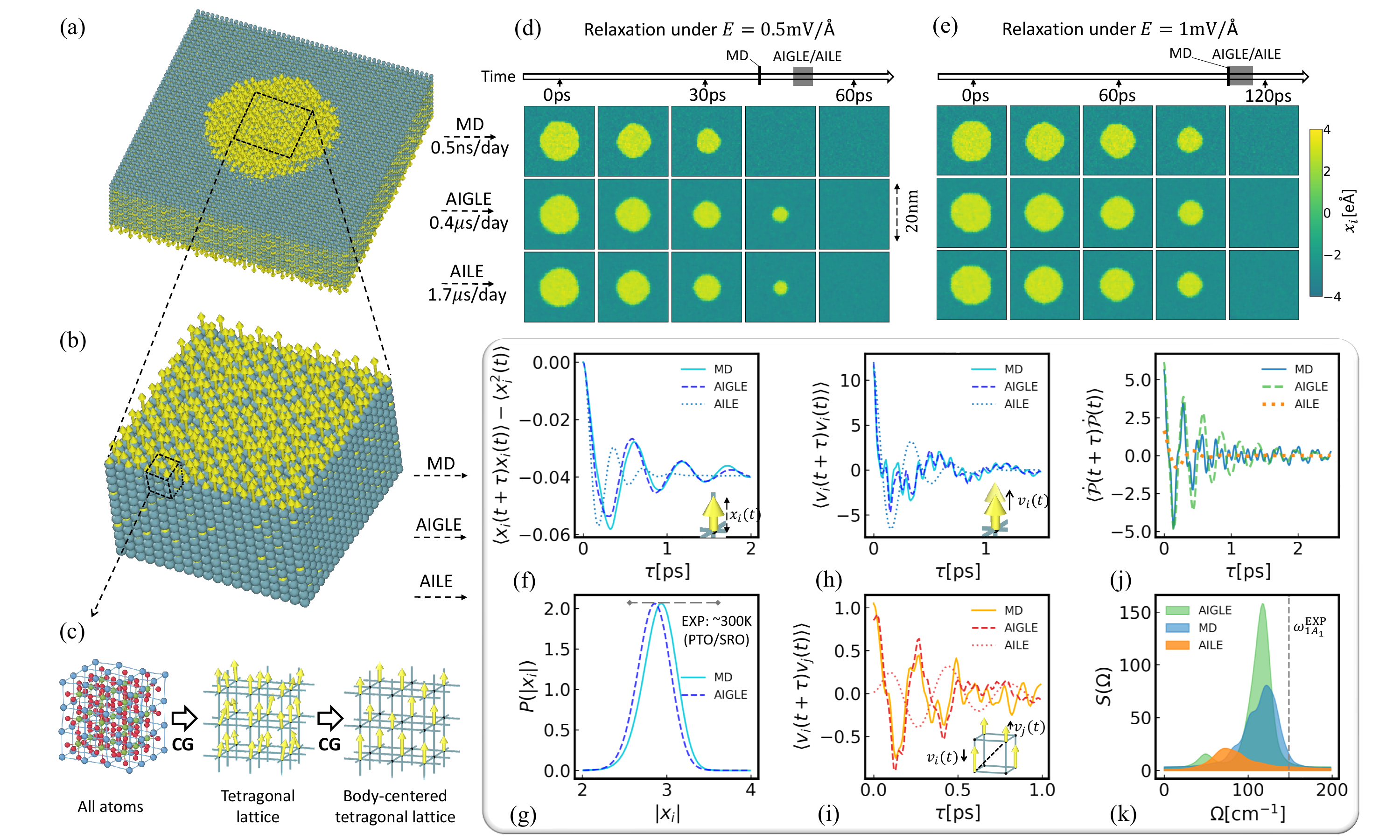}
    \caption{(a) A (non-equilibrium) configuration of \pto showing a cylindrical up-polarized domain in an environment of opposite polarization. Spheres are Ti atoms, Pb and O atoms are not shown. The arrows depict the local dipole moments assigned to Ti-centered elementary cells. (b) A configuration within the up-polarized domain. (c) Coarse-graining procedure. (d,e) Relaxation dynamics of the local dipoles in an atomic layer perpendicular to the direction $\bm{\hat{z}}$ of spontaneous polarization under electric fields $E=0.5$mV/$\ang$ (d), and $E=1$mV/$\ang$ (e). The three horizontal sequences of panels in (d) and (e) depict the evolution of the yellow domain from MD, AIGLE, and AILE. The pixels correspond to the $S$ sites. In the MD panels, they give the magnitude and sign of the microscopic dipoles $\{p_{jz}\}$, as per scale on the right. In the AIGLE and AILE panels the pixels associated with the $S_1$ sites give the CG dipoles $\{x_{j}\}$, while those associated with the $S_2$ sites are the average of the CG dipoles at the neighboring $S_1$ sites. The lifetimes of the cylindrical domain are shown on the two time axes. The solid vertical bars, at 40ps for $E=0.5$mV/A, and at 104ps for $E=1$mV/A, are extracted from two MD simulations. The grey rectangles are extracted from nine independent AIGLE and AILE simulations for each value of the electric field.
    (f) Shifted autocorrelation function (ACF) of the local dipole $x_i(t)$, in units of $\mathrm{(e \AA)}^2$, from MD, AIGLE, and AILE. (g) Equilibrium probability distribution $P(|x_i|)$ of the dipole magnitude from MD and AIGLE. The AILE result coincides with AIGLE and is not reported. The grey dashed line shows the range of the average dipole magnitude $\langle |x_i(t)| \rangle$ from different experiments~\cite{nishino2020,Dahl2009,Morita2004}.  (h) ACF of the time derivative, $v_i(t)$, of $x_i(t)$, in units of $\mathrm{(e \AA/ps)}^2$. (i) Cross-correlation function of $v_i(t)$ and $v_j(t)$ on adjacent sites in the $S_1$ lattice, in units of $\mathrm{(e \AA/ps)}^2$. (j) ACF of $\dot{\mathcal{P}}(t)$, the time derivative of the polarization $\mathcal{P}(t)$, in units of $\mathrm{(\mu C / cm^2 / ps)}^2$. (k) 
    Gaussian convoluted Fourier transform
    $S(\Omega)$ of the ACF of $\dot{\mathcal{P}}(t)$ (see text for details). The grey dashed line indicates the peak frequency $\omega^{\mathrm{EXP}}_{\mathrm{1A_1}}$ of the Raman spectroscopy feature associated to the zone-center $\mathrm{1A_1}$ transverse optical phonon~\cite{foster1993anharmonicity}.   }
    \label{fig:eqnoneq}
\end{figure*}

We run NVT-MD to generate the training data. The lattice parameters are fixed to $a=b=3.93\ang$ and $c\approx 4.04\ang$. For $E=0$, we run equilibrium NVT-MD in a $8\textrm{nm}\times 8\textrm{nm}\times5\textrm{nm}$ supercell comprising a single ferroelectric domain. The system is illustrated in Fig.~\ref{fig:eqnoneq} (b), where yellow arrows represent the local dipole moments. For $E=0.5$mV/A and $E=1$mV/A along +$z$, we run near equilibrium NVT-MD in a $20\textrm{nm}\times 20\textrm{nm}\times5\textrm{nm}$ supercell that initially contains two opposite ferroelectric domains, i.e., a nearly cylindrical up(+$z$)-polarized domain having a radius of about $6\textrm{nm}$, embedded in an environment of opposite polarization, as illustrated in Fig.~\ref{fig:eqnoneq}(a). To reduce the energy cost of the cylindrical interface
the up-polarized domain shrinks during the simulation, in spite of the applied field favoring up-polarization, with a longer relaxation time when $E$ is larger. In the MD simulations, we calculate the trajectories of all the local dipoles $\{\bm{p_j}(t)\}$. Taking the local dipoles in the tetragonal \pto lattice as CVs ( Fig.~\ref{fig:eqnoneq}(c, middle)), the degrees of freedom are one-fifth of the atom coordinates (Fig.~\ref{fig:eqnoneq}(c, left)). Further coarse-graining is motivated by the following considerations.
The time correlations of the Cartesian components of the local dipole velocities, i.e., $\langle \dot{p}_{j\alpha}(t_0+\tau)\dot{p}_{j\beta}(t_0)\rangle$  for $\alpha,\beta\in\{x,y,z\}$, indicate that the correlations for $\alpha\neq\beta$ are negligible compared to those for $\alpha=\beta$.
Thus, we can reduce by one-third the CVs by retaining only the $z$-components, $\{p_{jz}(t)\}$, of the local dipoles, which are related to spontaneous polarization. Nearest neighbor dipoles are strongly correlated, because the oxygen atoms, whose displacements contribute to the polarization the most, are shared between adjacent cells. As a consequence, further coarse-graining is possible by blocking into a single dipole pairs of nearest-neighbor dipoles of the original simple tetragonal lattice $S$. The blocking operation defines two interpenetrating body-centered tetragonal (BCT) lattices $S_1$ and $S_2$ obtained from $S$ by bipartition. We assume that our choice of CG dipoles corresponds to $S_1$, as illustrated in the right panel of Fig.~\ref{fig:eqnoneq}(c).  
If $\bm{a}=a\bm{\hat{x}}$, $\bm{b}=b\bm{\hat{y}}$, $\bm{c}=c\bm{\hat{z}}$ are the (conventional) unit cell vectors of $S$, the (conventional) unit cell vectors of $S_1$ are $\bm{u}=\bm{a}+\bm{b}$, $\bm{v}=-\bm{a}+\bm{b}$ and $\bm{w}=2\bm{c}$. Let $L$ and $A$ be the size and the adjacency matrix, respectively, of $S_1$.  Each site-$i$ of $S_1$ has 12 neighboring sites (in the sense of graph adjacency on $S_1$), displaced by ($\pm\bm{a}\pm\bm{b}$), $(\pm\bm{a}\pm\bm{c})$ and $(\pm\bm{b}\pm\bm{c})$, respectively. The corresponding CVs are denoted by  $\tilde{\bm{p}} = (p_{iz})_{i\in[1,L]}$. By construction, the degrees of freedom in $\tilde{\bm{p}}$ are one-30th of the atomic coordinates, but the polarization along $\bm{\hat{z}}$ is left unaffected. Then, we apply a truncated Gaussian filter in time to $\tilde{\bm{p}}(t)$ to remove high-frequency contributions. The resulting CVs are called 
$\{\bm{x}_{(n)}\}$:
\begin{equation}
\small
    \bm{x}_{(n)} =  \sum_{l=0}^{3\varsigma} \frac{\exp(-\frac{q^2}{2\varsigma^2})}{\sum_{l=0}^{3\varsigma} \exp(-\frac{q^2}{2\varsigma^2})}\tilde{\bm{p}}(t=n\Delta t - q\delta t).
\end{equation}
Here $\delta t=2$fs, $\Delta t=5\delta t$ and $\varsigma=40$, as in Sec.~\ref{sec-domain}. The polarization of the system is $\mathcal{P}(t=n\Delta t)=2 \sum_i x_{i,(n)} / V$.

The AIGLE model for $\{\bm{x}_{(n)}\}$ is:  
\begin{equation}\label{gle:cgld}
\small
\begin{split}
    M\bm{a}_{(n)}  
    =& -\nabla_{\bm{x}} G(\bm{x}_{(n)}) + pM\bm{E}_{(n)}  \\
    +& \sum_{l=0}^{n-1}  MK_{(l+\frac{1}{2})} \bm{v}_{(n-l-\frac{1}{2})} \Delta t + \bm{R}_{(n)},
\end{split}
\end{equation}
under local kernel approximation. We use for multi-dimensional AIGLE the same notation adopted in Eq.~(\ref{gle:domain}) for one-dimensional AIGLE. In principle, the external field $\bm{E}_{(n)}$ can vary in space and time, but we consider here only fields that are time-independent and uniform in space. For the free energy $G(\bm{x})$ we assume a simple polynomial form, $G(\bm{x})=\sum_{i} (b_1x_i^2+b_2x_i^4) + \sum_{ij}  A_{ij} b_{ij} (x_i-x_j)^2$, 
suggested by effective Hamiltonian models~\cite{zhong1995first}. By symmetry, $b_{ij}=b_3$ if the sites $i$ and $j$ are separated by ($\pm\bm{a}\pm\bm{b}$), and we set $b_{ij}=b_4$, otherwise. The $b$ coefficients are assumed to be independent of $E$, as appropriate in the linear response regime. Hence, we limit simulations to $E\leq 1$mV/A. Our model for $G$
is short-ranged but captures well the dipole-dipole interactions of the DP model within the cutoff radius of the latter. Long-range electrostatic interactions among the dipoles have a negligible effect on the ferroelectric transition 
in \pto~ (see, e.g., Ref.~\cite{xie2022ab} and references therein). 

Training proceeds through several steps. We first predetermine $G$ with equilibrium MD data ($E=0$) by force matching. Then, we calculate the memory kernel with the same data under local kernel approximation and train the multi-dimensional GAR model using $R_{(n)}$ as time series data. For the noise at site-$j$, the GAR model includes the noise history of site-$j$ and of its neighbors on $S_1$ displaced by ($\pm\bm{a}\pm\bm{c}$),($\pm\bm{b}\pm\bm{c}$) or ($\pm 2\bm{c}$).
In the last step, we retrain $G$ and $p$ with non-equilibrium MD data $(E>0)$. The details are in the SI. The corresponding AILE model is defined by the Markovian approximation of AIGLE as in 
the one-dimensional case.     

The relaxation dynamics of the cylindrical domain in Fig.~\ref{fig:eqnoneq}(a), under weak applied field, is illustrated in Fig.~\ref{fig:eqnoneq}(d), for $E=0.5$mV/A, and in Fig.~\ref{fig:eqnoneq}(e), for $E=1$mV/A. The noise in AIGLE and AILE trajectories is at the origin of the observed fluctuations in the domain lifetime. Within the uncertainty of the noise, AIGLE and AILE lifetimes coincide, suggesting that non-Markovian effects should be negligible. Indeed, domain shrinking is caused primarily by surface tension, which acts to reduce the area of the interface between domains, a systematic effect originating from  
the gradient of the free energy. The MD lifetime is deterministic and is extracted from a single trajectory. It agrees with AIGLE/AILE within the uncertainty of the noise for $E=1$mV/A (Fig.~\ref{fig:eqnoneq}(e)), but is approximately $10$ps shorter than AIGLE/AILE for $E=0.5$mV/A (Fig.~\ref{fig:eqnoneq}(d)). This discrepancy is likely due to the inaccuracy of the simple polynomial model adopted for the FES. Non-Markovian effects should be more pronounced for larger cylindrical domains, where the surface tension is smaller. Simulation of much larger domains would be feasible with AIGLE and AILE but not with all-atom MD, hampering direct comparison for these settings. The special case of a planar domain wall dynamics under applied field was considered in Sec.~\ref{sec-domain}, where it was found that non-Markovian effects play a role for very weak fields.

Next, we consider a uniformly polarized bulk sample in the absence of an external field ($E=0$). Static and dynamic properties of the dipoles are reported in Fig.~\ref{fig:eqnoneq}(f-k). Memory and noise effects are more pronounced in the equilibrium dynamics of the bulk than in the relaxation dynamics of a cylindrical interface.
Indeed, the MD ACF of an individual CG dipole $x_i$ is reproduced accurately by AIGLE but not by AILE (Fig.~\ref{fig:eqnoneq}(f)). At the same time, nearly identical results are obtained with AIGLE and AILE for static properties like the probability distribution $P(|x_i|)$ of the local dipole, reported in Fig.~\ref{fig:eqnoneq}(g), as expected from the fact that AIGLE and AILE yield the same equilibrium Boltzmann distribution. On the scale of Fig.~\ref{fig:eqnoneq}(g), AIGLE and AILE are identical and only AIGLE is reported.  The AIGLE distribution overlaps almost perfectly with the MD distribution barring a minor overall shift, much smaller than the range of the average dipole magnitudes extracted from experiments. 

The remaining panels in the figure confirm the importance of non-Markovian effects. Fig.~\ref{fig:eqnoneq}(h) shows that the ACF of $v_i(t)$, the time derivative of $x_i(t)$, is reproduced accurately by AIGLE but not by AILE. Also, the cross-correlation function between the time derivatives of neighboring dipoles shown in Fig.~\ref{fig:eqnoneq}(i) is reproduced well, at least up to about $0.5$ps, by AIGLE but not by AILE. These results suggest that the adopted local kernel approximation, which uses an optimized one-body memory kernel and many-body-correlated noise, can capture the short-range correlations among the dipoles that should dominate the fluctuation and dissipation of observables like the 
spontaneous polarization $\mathcal{P}$. Indeed, the ACF of $\dot{\mathcal{P}}(t)$, the time derivative of $\mathcal{P}(t)$, displayed in Fig.~\ref{fig:eqnoneq}(j), shows that AIGLE captures its dominant oscillatory frequency, while AILE misses it completely. However, at larger lagging times $\tau$ in the interval $[0.5,1.5]$ps AIGLE fails to reproduce the weak out-of-phase oscillations observed in MD. This behavior may originate from anharmonic couplings between vibrational modes that are not captured in the CG model. Neglect of long-range correlations in the noise could be another source of errors, as suggested by the observation that AIGLE would overestimate $\langle\dot{\mathcal{P}}^2(t)\rangle$ by about $30\%$ if the GAR model did not include the history dependence of neighbors separated by $(\pm 2 \bm{c})$. Thus, including longer-range correlations may improve the accuracy of the model. 
This may be possible by adopting
a more elaborate GAR model for the noise while retaining the simple local kernel approximation of AIGLE.
It is also instructive to compute $S(\Omega)$, the spectrum of $\langle \dot{\mathcal{P}}(t+\tau) \dot{\mathcal{P}}(t) \rangle$, which can be compared with experimental infrared spectroscopy. The spectra from MD, AIGLE, and AILE, given by $S(\Omega)=\Re \int_0^{\infty} d\tau\exp(-i\Omega \tau)\langle \dot{\mathcal{P}}(t+\tau)\dot{\mathcal{P}}(t) \rangle$, are reported in Fig.~\ref{fig:eqnoneq}(k), upon Gaussian broadening with full width at half maximum of $12\mathrm{cm}^{-1}$. 
As expected from the real-time data, the AILE peak in Fig.~\ref{fig:eqnoneq}(k) is significantly weaker than the other two, while AIGLE is stronger than MD, reflecting a sharper spectral feature. AIGLE reproduces well the peak frequency of MD, while AILE is red-shifted by approximately $40\mathrm{cm}^{-1}$. The spectral feature in Fig.~\ref{fig:eqnoneq}(k) is associated with the zone-center $\mathrm{1A_1}$ transverse optical phonon, which is both infrared and Raman active.
The corresponding feature from Raman scattering experiments lies at $\omega^{\mathrm{EXP}}_{\mathrm{1A_1}}=148.5\mathrm{cm}^{-1}$~\cite{foster1993anharmonicity}, with a full width at half maximum (FWHM) of approximately $30\mathrm{cm}^{-1}$,  while the MD FWHM is $43\mathrm{cm}^{-1}$ and that of AIGLE is $23\mathrm{cm}^{-1}$. 
The red shift of the MD/AIGLE peak at $120\mathrm{cm}^{-1}$, relative to the experiment, is mainly due to the adopted DFT approximation.

The above results show that AIGLE with the local kernel approximation can capture to a large extent the dynamic behavior of the CVs predicted by MD for bulk \pto. At the same time, AILE, while equivalent to AIGLE for static properties, can not capture dynamical correlations when memory is important.  

Finally, a comment on computational efficiency is in order. When modeling the dynamics of a $20\mathrm{nm}\times 20\mathrm{nm} \times 5\mathrm{nm}$ supercell on one Nvidia-A100 GPU, MD runs at $0.5$ns/day, AIGLE at $0.
4\mu$s/day, and AILE at $1.7\mu$s/day. Thus, the speedup over MD is of three orders of magnitude for both AIGLE and AILE. Moreover, AIGLE and AILE use significantly less memory than MD, facilitating simulations of significantly larger supercells.

\section{Discussion}\label{sec-outlook}

We have introduced a practical scheme to construct coarse-grained GLE models from MD trajectories. Our approach does not rely on the formal projection of MD onto the space of the CVs. As a consequence, the GLE construct is not exact, but should rather be viewed as a physically motivated approximation. While the idea of parameterizing GLE models with data extracted from MD trajectories dates back to at least 50 years~\cite{MCCOY1975431}, we exploit here modern techniques, such as machine learning and deep neural network representations, to generate extensive training data sets with MD and to construct the correlated noise model in the GLE. This enables us to construct AIGLE models, consistent with the microscopic dynamics, for one- and multi-dimensional CVs. Multi-dimensional AIGLE is not a trivial extension of its one-dimensional counterpart, and requires a local variational approximation for the memory kernel and a nearsightedness approximation for the correlated noise.  The latter could be formulated only for systems in which local CVs reside on sites with a fixed topology described by an adjacency matrix or a graph, such as crystals and individual polymeric molecules. How to extend the approach to more general disordered systems remains an open issue. Here, we considered mesoscale processes in \pto, a ferroelectric crystal, to illustrate the scheme and test its validity.                  
 
When used to study one-dimensional interface dynamics, AIGLE can model rare events on glassy landscapes caused by
nucleation and growth at the atomistic level, reproducing the interface evolution driven by a weak applied field at a much lower computational cost than MD. In contrast to MD, AIGLE can access very rare events, revealing that, in the ``slow'' creep regime, when the time scale of the events is much longer than that of the memory, the scaling law for the domain velocity may deviate significantly from that of the ``fast'' creep regime, due to non-Markovian effects. 
  

When applied to the dynamics of extensive CVs, AIGLE can model the relaxation of an elastic interface of any shape, a special case of extended defects, while still keeping the bulk dynamics of the CVs consistent with MD. These features distinguish AIGLE from other multi-scale models with more drastic levels of coarse-graining, such as, e.g., a Landau-Ginzburg field theory of the extensive CVs in the continuum limit. A field theory model can not provide atomistic level resolution of an interface, or correctly describe the vibrational spectrum of a global order parameter like the electric polarization at low but nonzero frequency. In ferroelectric materials polarization dynamics at low frequency is typically dominated by an optical phonon mode that cannot be reduced to white noise, and cannot be modeled by AILE. In this context, AIGLE captures many-body correlations between CV components that are topologically close when the distance is measured in a graph. This feature is the key difference between a truly multidimensional GLE and a set of one-dimensional GLEs with independent frictions and noises. 
 
Our study provides also examples where non-Markovian effects are irrelevant. In the CG lattice dynamics of \pto AIGLE and AILE give similar results for the motion of an interface dominated by a systematic driving force like the surface tension. In that case, memory effects are negligible, but
our study shows that they may become important for glassy dynamics.
It would be interesting to investigate the effect of driving fields that vary in space and time. Terahertz control of materials is an area of growing importance due to novel experimental developments~\cite{li2019terahertz}. For controlling fields within the frequency range of atomic/molecular vibrations, non-Markovian memory and noise effects could be in resonance with the external controlling field, coupling the latter to collective behavior associated with domain motion and/or phase transitions. 

Modeling CG lattice dynamics with AIGLE or AILE brings us to a scale where phenomena are typically treated with continuum models. These phenomena include general domain dynamics and phase separation in the condensed phase, which occur in ferromagnets
\cite{faghihi2013phase}, ferroelectrics~\cite{chen2008phase}, and alloys~\cite{allen1979microscopic, cahn1958free}. Other phenomena, important in the fabrication and characterization of nanomaterials, include morphology evolution in epitaxial growth~\cite{ momeni2018multiscale} and height fluctuations of two-dimensional membranes ~\cite{fasolino2007intrinsic}. In these contexts, the application of phase field models is very popular, whereby a continuum approximation is imposed {\it{ a priori}} and partial differential equations are constructed, guided by symmetry and physical intuition. This approach often captures the correct qualitative physics. However, when defects like impurities, grain boundaries, and domain walls are present, {\it ad hoc} continuum approximations fitted to few experimental observations, may be insufficient. When the role of defects is important, lattice models for the local dipole moments, local strains, and spins should be more reliable, as defect dynamics could be incorporated in lattice models by coupling homogeneous CVs on a lattice to a finite number of virtual particles representing mobile defects.
Along this line, one may be able to model notoriously difficult processes, such as those leading to the fatigue of ferroelectric devices when the dynamics of point defects gradually impacts the dynamics of domain walls over large space and time scales~\cite{dawber2005physics}.

All the applications discussed in the present work focused on the near-equilibrium regime, where the dynamics is constrained by the 2FDT. However, our methodology could be extended to far-from-equilibrium situations, where a governing principle like the 2FDT does not apply.
A regression-based approach like AIGLE can be adapted to deal with these situations, whereas conventional approaches based on ACFs would lose the convenience of direct construction of memory and noise terms. How to extend AIGLE to deal with far-from-equilibrium phenomena is a direction that we intend to explore in future studies.
  
\section*{Methods}
Here, we illustrate the learning procedure for the uni-variant GLE (Eqs.~(\ref{eom}-\ref{ag})). We will use $\mathcal{F}_{(n)}$ as an abbreviation for $-\nabla G(x_{(n)}) + F_{(n)}$. Also, without loss of generality, we assume $m=1$.

\subsection*{Separation of noise}
The first step of learning relies on equilibrated MD trajectories $\kappa=\{ x_{(n)}  | n\in [0,N]\}$ with ergodic fast degrees of freedom. $v_{ (n+\frac{1}{2})}$ and $a_{(n)}$ are computed from Eq.~(\ref{integrator}). $v_{ (n)}$ is further determined as the average of $v_{ (n+\frac{1}{2})}$ and $v_{ (n-\frac{1}{2})}$.
We turn the ensemble-averaged orthogonality condition $\langle R_{(n)} v_{(0)}\rangle=0$ to a time-averaged one. To achieve that, we introduce the shifted GLE with an arbitrary starting point $n_0 \geq 0$:
\begin{equation} \label{eq:s1}
    a_{(n)} = \mathcal{F}_{(n)} + \sum_{s=0}^{n-n_0-1}  K_{(s+\frac{1}{2})} v_{(n-\frac{1}{2}-s)} \Delta t + \tilde{R}^{(n_0)}_{(n)}.
\end{equation}
Here, $\tilde{R}^{(n_0)}_{(n)}$ is a shifted noise and $n > n_0$ is required. As demonstrated in Ref.~\cite{kubo1966fluctuation},  $\langle \tilde{R}^{(n_0)}_{(n_0+k)}\tilde{R}^{(n_0)}_{(n_0)}\rangle=\langle R_{(n +k)}R_{(n )}\rangle$ for a stationary noise series when $n\rightarrow \infty$.

For a given CV trajectory, the shifted noise $\tilde{R}^{(n_0)}_{(n)}$ is explicitly computed by inverting Eq.~(\ref{eq:s1}):
\begin{equation}\label{shifted}
    \tilde{R}^{(n_0)}_{(n)} =
    a_{(n)} - \mathcal{F}_{(n)} - \sum_{s=0}^{n-n_0-1}  K_{(s+\frac{1}{2})} v_{(n-s-\frac{1}{2})} \Delta t.
\end{equation}
For $n=n_0$, we let $\tilde{R}^{(n)}_{(n)} =  a_{(n)} - \mathcal{F}_{(n)}$.
The time-averaged estimator of $\langle R_{(k)} v_{(0)}\rangle$  can be written as $\zeta_k = \frac{1}{N_k}\sum_{n_0=0}^{N_k-1} \tilde{R}^{(n_0)}_{(n_0+k)}v_{(n_0)}$,
where $N_k=N-k$. Note that $\zeta_0$ only depends on the force fields while $\zeta_k$ also depends on the memory kernel for $k>0$. It is not recommended, for numerical stability, to train $\mathcal{F}$ by imposing the orthogonality condition $\zeta_0=0$ directly. We recommended, instead, to train $\mathcal{F}$ by minimizing the noise within a maximum-likelihood perspective, and further decouple the training of $\mathcal{F}$ and $\mathcal{K}$ for stability and efficiency. To achieve these goals, we first define the constrained optimization problem:
\begin{equation}\label{optimization}
\begin{split}
     \underset{\mathcal{F}^\theta, K^\theta}{\text{minimize}}  \ & \underset{\kappa\sim\pi_\kappa}{\mathbb{E}}\sum_{n=m_K}^{N-1} |\tilde{R}^{(0)}_{(n)}|^2  \\
 \text{subject to} \ \  & \underset{\kappa\sim\pi_\kappa}{\mathbb{E}}\zeta_k  = 0, \ \ k\in[1,m_K].
\end{split}
\end{equation}
Here, $\pi_\kappa$ denotes an ensemble of $\kappa$ trajectories.
The ensemble average $\mathbb{E}_{\kappa\sim\pi_\kappa}$ is not necessary when $\kappa$ is ergodic and sufficiently long. But in practice averaging over multiple finite-size trajectories is preferred.
$m_K$ is the finite memory cutoff of $K$. $\mathcal{F}^\theta$ and  $K^\theta$ are the parameters of $\mathcal{F}$ and $K$, respectively. $\mathcal{F}$ can be any differentiable parameterized function, including neural networks. Eq.~(\ref{optimization}) should be transformed into an unconstrained problem in practical applications. Notice that the constraint $\mathbb{E}_{\kappa\sim\pi_\kappa}\zeta_k = 0$ can be written equivalently  as 
\begin{equation}\label{aveq}
\begin{split}
       \underset{\kappa\sim\pi_\kappa}{\mathbb{E}} \sum_{n_0=0}^{N_k-1} & ( a_{(n_0+k)} -\mathcal{F}_{(n_0+k)} ) v_{(n_0)}  \\
    &=  \underset{\kappa\sim\pi_\kappa}{\mathbb{E}} \sum_{s=0}^{k-1}   K_{(s+\frac{1}{2})}\Delta t    \sum_{n_0=0}^{N_k-1}v_{(n_0+k-s-\frac{1}{2})} v_{(n_0)} .
\end{split}
\end{equation}
Considering $k\in[1,m_K]$, Eq.~(\ref{aveq}) can be written in matrix form as $\mathcal{Y}=\mathcal{C}\mathcal{K}$. $\mathcal{Y}$ and $\mathcal{K}$ are vectors of length $m_K$. $\mathcal{C}$ is a $m_K\times m_K$ lower triangular matrix. The left-hand side of Eq.~(\ref{aveq}) is the $k$-th entry of $\mathcal{Y}$.  The $j$-th entry of $\mathcal{K}$ is $\mathcal{K}_j=K_{(j-\frac{1}{2})}\Delta t$. And $\mathcal{C}_{kj}=\mathbb{E}_{\kappa\sim\pi_\kappa}\sum_{n_0=0}^{N_k-1}v_{(n_0+k-j+\frac{1}{2})} v_{(n_0)} $ when $m_K\geq k\geq j \geq 1$. Hence, the least-square solution of Eq.~(\ref{aveq}) can be written as $\mathcal{K} = \text{Inv}(\mathcal{C}^T\mathcal{C})\mathcal{C}^T\mathcal{Y}$.  $\text{Inv}$ is the pseudo-inverse operator computed from single-value decomposition with a cutoff ratio to avoid numerical instability.  

We are then able to approach the solution of Eq.~(\ref{optimization}) practically by interleaving $n^{\mathrm{GD}} \geq 1$ unconstrained optimization steps towards 
\begin{equation} 
     \underset{\mathcal{F}^\theta }{\text{minimize}}  \underset{\kappa\sim\pi_\kappa}{\mathbb{E}}\sum_{n=m_K}^{N-1} |\tilde{R}^{(0)}_{(n)}|^2 
\end{equation}
with one iteration of 
\begin{equation}\label{scf-gle}
     \begin{split}
     \mathcal{K} &\rightarrow (1-\epsilon)\mathcal{K} + \epsilon  \text{Inv}(\mathcal{C}^T\mathcal{C})\mathcal{C}^T\mathcal{Y}, \\
     \mathcal{K} &\rightarrow \mathcal{K} -  \mathcal{K}_{m_K} \bm{1}_{m_K}.
     \end{split}
\end{equation}
The parameter $\epsilon\in(0,1)$ should be small enough for stability. In this work we use $\epsilon=0.01$.
The second step in Eq.~(\ref{scf-gle}) forces $\mathcal{K}_{m_K}=0$ over the course of training.

\subsection*{Training of the GAR model}
In the previous step, the noise $R_{(n)}=\tilde{R}^{(0)}_{(n)}$ is extracted from $a_{(n)}$. Then one can establish a GAR model with $R_{(n)}$ as data. The GAR parameters include the linear coefficients $\phi=(\phi_{(1)},\cdots, \phi_{(k)})$ and the parameters  $\{\mu^\theta,\sigma^\theta\}$ of the neural network.
We define the maximum likelihood loss function
\begin{equation}
\small
    \mathcal{L}^{\mathrm{GAR}} = \underset{\kappa\sim\pi_\kappa}{\mathbb{E}}\sum_n \ln \sigma_{(n)}^2 + \frac{(R_{(n)}-\sum_{k=1}^{m_A} \phi_{(k)}R_{(n-k)}-\mu_{(n)})^2}{\sigma_{(n)}^2}.
\end{equation}
It is not recommended to minimize $\mathcal{L}^{\mathrm{GAR}}$ directly with respect to all the parameters without constraints. Overfitting the data should be avoided for the long-term stationarity of the GAR model. This is crucially important for simulating AIGLE at or above the $\mu$s scale, much longer than the picosecond/nanosecond duration of the MD trajectories. So, we harness the GAR model by imposing on $\phi$ the constraint that they should satisfy the Yule-Walker equation. For a given CV trajectory, let $\lambda_{(k)}$ be the estimator of the noise ACF, given by  $\lambda_{(k)}=\frac{1}{N_k-m_A}\sum_{n=m_A}^{N_k-1} R_{(n+k)}R_{(n)}$. Let the vector $\Lambda$ be $\Lambda = \underset{\kappa\sim\pi_\kappa}{\mathbb{E}}(\lambda_{(1)}, \cdots,\lambda_{(m_A)})$. Let the $m_A\times m_A$ matrix $\mathcal{M}$ be  $\mathcal{M}_{jk}=\underset{\kappa\sim\pi_\kappa}{\mathbb{E}}\lambda_{(|j-k|)}$. The Yule-Walker equation for a standard AR($m_A$) model is $\Lambda=\mathcal{M}\phi$, the least square solution of which can be written as $\phi^{\mathrm{YW}}= \text{Inv}(\mathcal{M}^T\mathcal{M})\mathcal{M}^T\Lambda$. Using the Yule-Walker solution as a constraint, we optimize the GAR model by interleaving $n^{\mathrm{GD}}$ unconstrained optimization steps towards
\begin{equation}
         \underset{ \mu^\theta,\sigma^\theta }{\text{minimize}}\ \   \mathcal{L}^{\mathrm{GAR}}
\end{equation}
with one iteration of 
\begin{equation}\label{scf-gar}
         \phi  \rightarrow (1-\epsilon)\phi + \epsilon  \text{Inv}(\mathcal{M}^T\mathcal{M})\mathcal{M}^T\Lambda.
\end{equation}

Although in the formal presentation, the training of GAR is done after the training of the first step, in practice one can train GAR on the fly to simplify the implementation. 

\subsection*{Incorporation of near-equilibrium data}
In this step, we deal with additional datasets that violate detailed balance. 
We fix the memory kernel and the GAR model obtained for thermal equilibrium, assuming that they are approximately the same in near-equilibrium situations. 
The optimization task is simply
\begin{equation}\label{opt1}
     \underset{\mathcal{F}^\theta }{\text{minimize}}  \underset{\kappa\sim\pi_\kappa}{\mathbb{E}}\sum_{n=m_K}^{N-1} |\tilde{R}^{(0)}_{(n)}|^2 
\end{equation}
for the extended dataset. Here $\mathcal{F}^\theta$ may include the parameters of the external driving forces.

\subsection*{Data and Code availability}
The DP model and a minimal implementation of AIGLE are publicly available on Github~\cite{code}.

\section*{Acknowledgement}
    We thank Yucheng Yang, Huan Lei, William M Jacobs, Yixiao Chen, and Linfeng Zhang for fruitful discussions. All authors were supported by the Computational Chemical Sciences Center: Chemistry in Solution and at Interfaces (CSI) funded by DOE Award DE-SC0019394. 
    P.X. and W.E were also supported by a gift from iFlytek to Princeton University. P.X. was also supported by the Azure cloud computing mini-grant from the Center for Statistics and Machine Learning of Princeton University.  W.E was supported by the Basic Science Center of National Natural Science Foundation of China with Award NSFC No.12288101.
    The authors are pleased to acknowledge that the work reported in this paper was performed largely using the Princeton Research Computing resources at Princeton University which is a consortium of groups led by the Princeton Institute for Computational Science and Engineering (PICSciE) and the Office of Information Technology's Research Computing. This research also used resources of the National Energy Research Scientific Computing Center (NERSC), a U.S. Department of Energy Office of Science User Facility located at Lawrence Berkeley National Laboratory, operated under Contract No. DE-AC02-05CH11231 using NERSC award ERCAP0021510.



\renewcommand*{\bibfont}{\scriptsize}
\bibliography{pnas-sample}

\end{document}


\title{Supporting information: Ab Initio  Generalized Langevin Equation}
\author{Pinchen Xie, Roberto Car, Weinan E}
\maketitle



\makeatletter
\renewcommand{\theequation}{S\arabic{equation}}
\renewcommand{\thefigure}{S\arabic{figure}}
\renewcommand{\bibnumfmt}[1]{[S#1]}
\renewcommand{\citenumfont}[1]{S#1}

\section{Validation with a solvable model}\label{sec:validation}
We study a toy model with a known memory kernel~\cite{Adelman1974} in the MZ framework under thermal equilibrium conditions. The model consists of a one-dimensional harmonic chain of $N+1$ identical particles of mass $m$. The chain has one free and one fixed end. 
Let $\{z_i\}_{i\in[0,N]}$ and $\{p_i\}_{i\in[0,N]}$ be the coordinate and momentum of the particles in the harmonic chain. Let $z_0=0$ and $p_0=0$.
The Hamiltonian of the system is
\begin{equation}
    H(\{p_i\}, \{z_i\})=\sum_{i=0}^{N-1} \frac{p_{i+1}^2}{2m} + \frac{1}{2}\tilde{\omega}^2 (z_i-z_{i+1}-a)^2.
\end{equation}
  
The parameters of the system we simulate are $N=100$, $m=k_BT=\tilde{\omega}=1$. $x_N$ is connected to a Langevin thermostat with damping time $t_d=10$, representing a gas/solid interface. 
The CV $x$ is chosen to be the displacement of the free end of the chain from its equilibrium position, i.e., $x=x_N$.  
We use a MD time step $\delta t=0.1$ and obtain 200 trajectories of $x$, each lasting $t_{MD}=2000$ after initial equilibration. The trajectories are further coarse-grained with a time step $\Delta t = 4\delta t$ by sampling the MD data every 4 MD time steps.
The GLE ansatz for this system can be written as
\begin{equation}\label{gle:chain}
    a_{(n)} = -\omega_0^2x_{(n)} + \sum_{s=0}^{n-1}  K_{(s+\frac{1}{2})} v_{(n-s-\frac{1}{2})} \Delta t + R_{(n)}.
\end{equation} 
The neural network part of the corresponding GAR model is a feed-forward neural network with two hidden layers (size=10). The neural network only outputs $\mu_{(n)}$ in the GAR model. The $\sigma_{(n)}$ in the GAR model is parameterized directly by one scalar variable since its history dependence is unnecessary for this case study. The finite memory cutoff is taken to be $m_K=50$ for the memory kernel and $m_A=25$ for the GAR model.   The training is done with the Adam optimizer implemented in PyTorch~\cite{paszke2017automatic} with the default setting and learning rate of 0.001. Convergence is reached with approximately 3000 iterations. For each iteration, 50 trajectories of the CV are randomly chosen for optimization. The parameters $\epsilon$ for the iterative update of memory kernel and the Yule-Walker coefficient are all $0.01$. The interval $n^{\mathrm{GD}}$ is 10. The optimized $\omega_0$ is $0.09$. The optimized standard deviation of the white noise is $\sigma_{(n)}=0.721$. The colored noise $R$ and the white noise $w$ are both distributed normally on the dataset with negligible off-centering. Non-stationarity is not detected numerically for the GAR model.

\begin{figure}[thb]
\centering
\includegraphics[width=0.6\linewidth]{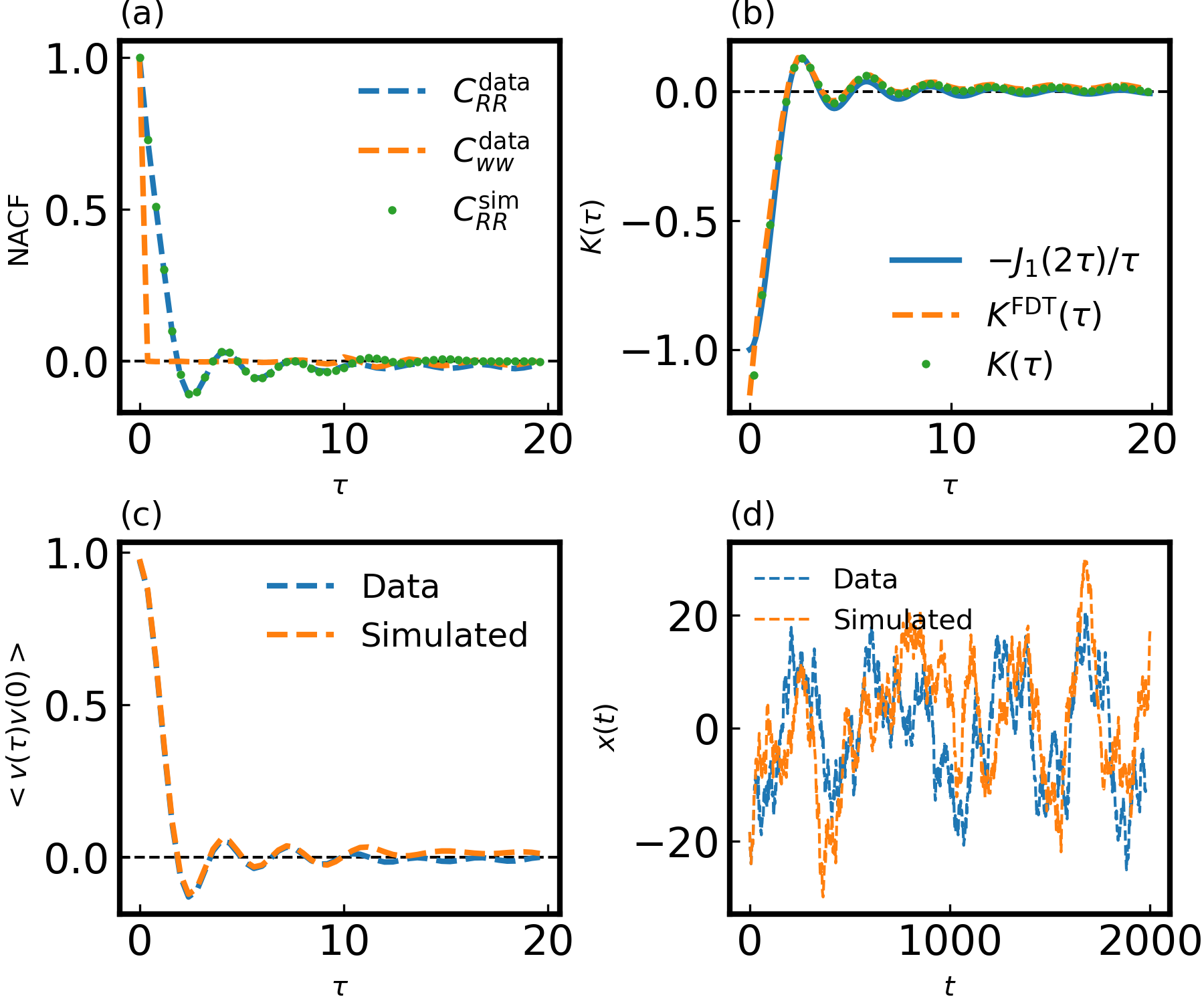}
\caption{(a) NACF of the noise. $C^{\text{data}}_{RR}(\tau)=\langle R(t_0+\tau)R(t_0)\rangle/\langle R(t_0)^2\rangle$ and   $C^{\text{sim}}_{RR}(\tau)=\langle R(t_0+\tau)R(t_0)\rangle/\langle R(t_0)^2\rangle$ are averaged over the dataset and the simulated trajectories, respectively. $C^{\text{data}}_{ww}(\tau)=\langle w(t_0+\tau)w(t_0)\rangle/\langle w(t_0)^2\rangle$ is the NACF of the residual noise from the dataset. (b) The optimized memory kernel $K(\tau)$ compared to the theoretical one and the 2FDT prediction $K^{\mathrm{FDT}}(\tau=n\Delta t)=-\beta\langle R_{(n)} R_{(0)}\rangle$. 
(c) The velocity autocorrelation function. (d) Comparison between an MD trajectory and a GLE trajectory with the same initial conditions.} 
\label{fig:chain}
\end{figure}

The outcome of the training is reported in Fig.~\ref{fig:chain}. The GAR model successfully reduces $w$ to almost white noise with delta-like normalized autocorrelation function (NACF), as shown in Fig.~\ref{fig:chain}(a). An obvious serial correlation would be present if we used an AR model instead. Both $R$ and $w$ are normally distributed on the dataset.  Fig.~\ref{fig:chain}(b) shows that the optimized $K(\tau)$ obeys 2FDT and agrees well with the theoretical result for an isolated infinite chain, $K^{\mathrm{MZ}}(\tau)=-J_1(2\tau)/\tau$, where $J_1$ is a Bessel function of the first kind. The discrepancy near $\tau=0$ is due to the inclusion of white noise from the Langevin thermostat. 
Fig.~\ref{fig:chain}(c) compares the velocity autocorrelation from the MD dataset and the AIGLE simulation. They display minor disagreement for $\tau > 10$. This is the error caused by substituting $w_{(n)}$ with the ideal white noise generator when propagating AIGLE. The same effect can be seen in Fig.~\ref{fig:chain}(a) on the slight discrepancy between $C^{\text{data}}_{RR}$ and $C^{\text{sim}}_{RR}$.
We also compare a simulated AIGLE trajectory and an MD trajectory with the same initial conditions in Fig.~\ref{fig:chain}(d). They display statistically equivalent behavior of stochastic harmonic oscillators. 

\section{ Continuous-time on-lattice GLE }\label{sec:cgle}
\subsection{Variational approach for memory kernel}
Following the notation used in the main text, the GLE for $\xx_i=(x_{i1}, \cdots,x_{id_s})^T$ is
\begin{equation}\label{app:gle}
\small
    M\bm{a}(t) = \bm{\mathcal{F}}(t)  + \int_0^t ds MK(s)\bm{v}(t-s) + \bm{R}(t).
\end{equation}
We recall the orthogonality condition $\langle \bm{R}(t) \bm{v}^T(0)\rangle=0$, and the 2FDT 
\begin{equation}
\langle \bm{v}(0)\bm{v}^T(0) \rangle K^T(s) = -  \langle (M^{-1}\bm{R})(0) (M^{-1}\bm{R})^T(s)\rangle,   
\end{equation}
for the equilibrium ensemble. 
We start by formulating a variational principle for the most general memory kernel, i.e., a kernel $K(s)=\{K_{i\alpha}^{j\beta}(s)\}$ represented by a dense tensor. Using Eq.~(\ref{app:gle}), the orthogonality tensor defined by $\Omega_{i\alpha}^{j\beta}(t)=\langle R_{i\alpha}(t)v_{j\beta}(0)\rangle$, can be written as
\begin{equation}
    \Omega_{i\alpha}^{j\beta}(t) = Q_{i\alpha}^{j\beta}(t)  - \int_0^t ds (MK)_{i\alpha}^{k\gamma}(s) C_{k\gamma}^{j\beta}(t-s),
\end{equation}
where the Einstein notation for repeated indices is used, $Q_{i\alpha}^{j\beta}(t)=\langle (Ma- \mathcal{F})_{i\alpha}(t) v_{j\beta}(0) \rangle$, and $C_{k\gamma}^{j\beta}(t-s)=\langle v_{k\gamma}(t-s) v_{j\beta}(0)\Theta(t-s)\rangle $, with $\Theta$ the Heaviside step function ($\Theta(x)=1$ for $x>0$, $\Theta(x)=0$ otherwise). In general, the deterministic force $\mathcal{F}$ can be defined by force matching. Assuming that the exact $\mathcal{F}$ is known, the diagonal element of the orthogonality tensor at $t=0$ vanishes, i.e., $\Omega_{i\alpha}^{i\alpha}(0)=Q_{i\alpha}^{i\alpha}(0)=0$

Given a canonical ensemble of trajectories
and the corresponding exact $\mathcal{F}$, the orthogonality condition implies that the exact memory kernel $K$ minimizes the canonical orthogonality loss functional $\tilde{\mathcal{L}}[K]$, defined by  
\begin{equation}
\tilde{\mathcal{L}}[K]=   \int_0^{\infty} \sum_{i,j,\alpha,\beta} |\Omega_{i\alpha}^{j\beta}(t)|^2  dt. 
\end{equation}
By exploiting locality in space and finite memory time, we approximate  $\tilde{\mathcal{L}}$ with $\mathcal{L}$ given by
\begin{equation}\label{lossfunc}
    \mathcal{L} [K](t_K)=\int_0^{t_K} \sum_{i,j,\alpha,\beta} |\Omega_{i\alpha}^{j\beta}(t)|^2 \Theta(d_c-d_{ij}) dt,
\end{equation}
where $t_K$ is the cutoff time of the memory, $d_c$ is the cutoff length of the correlation in space, and $d_{ij}$ is the graph geodesic distance between sites $i$ and $j$, given the adjacency matrix $A$ of the lattice. We require $K(s)=0$ for $s>t_K$.
A bulk crystalline system, under equilibrium conditions, has translational symmetry, and the complexity of evaluating $\mathcal{L} [K](t_K)$ is limited by $d_c$ and $t_K$, regardless of the size of the lattice. Then, the optimal $K=K^*$ for given $d_c$ and $t_K$ is found by minimization, i.e.,
\begin{equation}
    K^*=\mathrm{argmin}_K \mathcal{L} [K](t_K)
\end{equation}

\subsection{Local kernel approximation}
The approximation introduced in the previous subsection is still excessively complicated for use in practical applications. Here, we make a further important simplification, valid when the velocity correlations of the CVs in the extensive coarse-grained lattice dynamics are negligible beyond nearest neighbors. Then, $1<d_c<2$, and the functional in Eq.~(\ref{lossfunc}) becomes 
\begin{equation}\label{loss-local}
    \mathcal{L} [K](t_K)=\int_0^{t_K} \sum_{i,j,\alpha,\beta} |\Omega_{i\alpha}^{j\beta}(t)|^2 (\delta_{ij}+A_{ij}) dt.
\end{equation}
We also assume that $M_{i\alpha}^{j\beta}=\delta_{ij}\delta_{\alpha\beta}M_{\alpha}$ has been determined from the data by applying the equipartition theorem to the kinetic energy of the CVs. 
The local kernel approximation for $K$, adopted in the main text, assumes that only the on-site elements $K_{i\alpha}^{i\beta}(s)$ can be nonzero.
Intuitively, one expects that the main contribution to memory should be on-site. In practice, the local kernel approximation is justified when
the corresponding coarse-grained dynamics reproduces with sufficient accuracy the dynamics inferred from the underlying microscopic Hamiltonian. We notice that the local kernel approximation would be compatible with a less restrictive choice of the spatial cutoff $d_c$ in Eq.~(\ref{lossfunc}). In this way, the memory kernel will still be on-site but will depend effectively on longer range correlations between the CVs. 
Since, by translational symmetry,  $K_{i\alpha}^{i\beta}(s)$ does not depend on the site index $i$, we drop this index in the following, and use the notation $K_\alpha^\beta(s)$ for $K$, and $\Omega^{0}_{\alpha,\beta}$ for $\Omega_{i\alpha}^{i\beta}$
For the same reason, $\Omega_{i\alpha}^{j\beta}$ should depend only on the relative displacement  $\bm{\eta}_{ij}$ connecting sites $i$ and $j$ in the lattice.  
Let $\{ \bm{\eta}_{\iota}| \iota=1,\cdots,d_{\mathrm{nn}}\}$ be the set of different lattice vectors connecting nearest neighbor sites, and let $\eta_0$ be the null vector. Using the label $i_{[\iota]}$ to indicate a site  displaced from site-$i$ by $\bm{\eta}_{\iota}$, we further simplify the notation for  $\Omega_{i\alpha}^{i_{[\iota]}\beta}$ to $\Omega^{\iota}_{\alpha,\beta}$.
Then, for arbitrary $i$ and $j$, we have $\Omega_{i\alpha}^{j\beta}(\delta_{ij}+A_{ij})\in \{ \Omega^{\iota}_{\alpha,\beta}| \iota=0,\cdots,d_{\mathrm{nn}}\}$.
Then, Eq.~(\ref{loss-local}) becomes
\begin{equation}\label{loss-local2}
    \mathcal{L} [K](t_K)= L \int_0^{t_K} \sum_{\alpha,\beta} \sum_{\iota=0}^{d_{\mathrm{nn}}}|\Omega^{\iota}_{\alpha,\beta}(t)|^2 dt.  
\end{equation}
Here, 
$\Omega^{\iota}_{\alpha,\beta}(t)$ is given by 
\begin{equation}\label{omega-first}
    \Omega^{\iota}_{\alpha,\beta}(t)= Q^{\iota}_{\alpha,\beta}(t)  - \int_0^t ds M_{\alpha} \sum_{\gamma}K_{\alpha}^{\gamma}(s) C^{\iota}_{\gamma,\beta}(t-s).
\end{equation}
In Eq.~(\ref{omega-first}),
$Q^{\iota}_{\alpha,\beta}(t)= \frac{1}{L}\sum_i\langle (Ma- \mathcal{F})_{i\alpha}(t) v_{i_{[\iota]}\beta}(0) \rangle$,
and 
$C^{\iota}_{\gamma,\beta}(t-s)=\frac{1}{L}\sum_i\langle v_{i\gamma}(t-s) v_{i_{[\iota]}\beta}(0)\Theta(t-s)\rangle$ are directly calculated from the data. Writing this equation in matrix form (with respect to the indices $\alpha$ and $\beta$) leads to
\begin{equation}\label{omega-mat}
    \Omega^{\iota}(t) = Q^{\iota}(t)  - \int_0^t ds MK(s) C^{\iota}(t-s).
\end{equation}
Here all matrices are of size $d_s\times d_s$.
Taking the functional derivative of $\mathcal{L} [K](t_K)$ with respect to $M_\alpha K_{\alpha}^{\gamma}(s)$ one obtains 
\begin{equation}\label{fderiv}
    \frac{\delta \mathcal{L} [K](t_K)}{\delta M_\alpha K_{\alpha}^{\gamma}}(s) 
    -2L\int_s^{t_K} dt \sum_\beta \sum_{\iota=0}^{d_{\mathrm{nn}}} 
    \Omega_{\alpha,\beta}^\iota(t)C^\iota_{\gamma,\beta}(t-s)  
\end{equation}
Putting Eq.~(\ref{fderiv}) in matrix form and inserting Eq.~(\ref{omega-mat}) leads to
\begin{equation}
\small
\begin{split}
        \frac{\delta \mathcal{L} [K](t_K)}{\delta (MK)}(s) 
    &=-2L\sum_{\iota=0}^{d_{\mathrm{nn}}}\int_s^{t_K} dt   
      \Big(Q^{\iota}(t)(C^\iota)^T(t-s)  \\
      &- \int_0^t dr MK(r) C^{\iota}(t-r)(C^\iota)^T(t-s)\Big) \\
      &=
      -2L\sum_{\iota=0}^{d_{\mathrm{nn}}}\int_0^{t_K} dt   
      \Big(Q^{\iota}(t)(C^\iota)^T(t-s)  \\
      &- \int_0^{t_K} dr MK(r) C^{\iota}(t-r)(C^\iota)^T(t-s)\Big)
\end{split}
\end{equation}
The second equality holds because by definition $C^\iota(x)=0$ for $x<0$.  

Finally, $K^*$ is the least square solution of $\frac{\delta \mathcal{L} [K](t_K)}{\delta (MK)}(s)=0$, i.e.,
\begin{equation}\label{ls1}
\begin{split}
    \sum_{\iota=0}^{d_{\mathrm{nn}}} &\int_0^{t_K} dr MK(r) \int_0^{t_K} dtC^{\iota}(t-r)(C^\iota)^T(t-s)\\
    &=\sum_{\iota=0}^{d_{\mathrm{nn}}}\int_0^{t_K} dt   
       Q^{\iota}(t)(C^\iota)^T(t-s).
\end{split}
\end{equation}
In the continuous-time formulation of the GLE it is not evident how one could find the solution $K^*$ via a standard least-square algorithm. This will become evident in the next section where we use the discrete-time formulation of the GLE recommended in practical implementations.
\section{ Discrete-time on-lattice AIGLE} \label{sec:dgle}
\subsection{Formulation}\label{sec:dgle:formulation}
The discrete-time GLE is an approximation to the continuous-time GLE. Here we follow the same leap-frog strategy adopted for univariant AIGLE.
The discrete-time on-lattice AIGLE takes the form
\begin{equation} 
    M\bm{a}_{(n)}  = \bm{\mathcal{F}}_{(n)} + \sum_{l=0}^{n-1}  MK_{(l+\frac{1}{2})} \bm{v}_{(n-l-\frac{1}{2})} \Delta t + \bm{R}_{(n)}.
\end{equation}
Here $\Delta t$ is the time step of integration, $n$ is an integer, and $t=n\Delta t$ is the current time. 
We use $f_{(n)}$ as an abbreviation for any time-dependent function $f(n\Delta t)$. 
The integration scheme for the GLE is given by
\begin{equation}\label{integrator}
\begin{split}
    \bm{v}_{(n+\frac{1}{2})} & = \bm{v}_{(n-\frac{1}{2})} + \bm{a}_{(n)}\Delta t, \\
    \bm{x}_{(n+1)} &=\bm{x}_{(n)} + \bm{v}_{(n+\frac{1}{2})}\Delta t.
\end{split}
\end{equation}
We call $\{n\Delta t | n\in\mathbb{Z}  \}$ the integer grid and $\{(n-\tfrac{1}{2})\Delta t | n\in\mathbb{Z}  \}$ the shifted grid.  The trajectory data $x_{(n)}$ are given on the integer grid.
$\bm{v}_{(n-\frac{1}{2})}$ and $\bm{a}_{(n)}$ can are computed directly from Eq.~(\ref{integrator}). $\bm{v}_{(n)}$ on the integer grid is subsequently intepolated as $\bm{v}_{(n)}=(\bm{v}_{(n+\frac{1}{2})}+\bm{v}_{(n-\frac{1}{2})})/2$.

Let $t_K=m_K\Delta t$ be the cutoff of the memory time. Then, $K_{(l+\frac{1}{2})}=0$ for $l\geq m_K$. The quantity $Q^{\iota}_{\alpha,\beta}(n\Delta t)$ of the continuous-time formulation becomes
$Q^{\iota}_{\alpha,\beta, (n)}= \frac{1}{L}\sum_i\langle (Ma- \mathcal{F})_{i\alpha,(n)} v_{i_{[\iota]}\beta, (0)} \rangle$ in the discrete time representation, and
$C^{\iota}_{\gamma,\beta}((n-l+\tfrac{1}{2})\Delta t)$ becomes
$C^{\iota}_{\gamma,\beta, (n-l)}=\frac{1}{L}\sum_i\langle v_{i\gamma,(n-l+\frac{1}{2})} v_{i_{[\iota]}\beta,(0)}\Theta(n-l+\tfrac{1}{2})\rangle.$ 
To avoid messy indexing, in the following we make all the indices labelling a local order parameter component implicit. With this simplified notation $Q^{\iota}_{(n )}$ stands for the $d_s\times d_s$ matrix of which the $(\alpha,\beta)$-entry is $Q^{\iota}_{\alpha,\beta, (n )}$,   $C^{\iota}_{(n-l)}$ stands for the $d_s\times d_s$ matrix of which the $(\alpha,\beta)$-entry is $C^{\iota}_{\alpha,\beta, (n-l)}$, 
$K_{(l+\frac{1}{2})}$ stands for the $d_s\times d_s$ matrix of which the $(\alpha,\beta)$-entry is $K_{\alpha,(l+\frac{1}{2})}^\beta$.


\subsection{Memory kernel}
The results of Sec.~\ref{sec:cgle} can be translated straightforwardly into discrete-time notation. Thus, we go directly to the local kernel approximation and show how the optimal memory kernel can be obtained in practice. 
 
Eq.~(\ref{loss-local2}) becomes
\begin{equation}\label{loss-discrete}
    \mathcal{L} [K](t_K)=L \sum_{n=1}^{m_K} \sum_{\alpha,\beta} \sum_{\iota=0}^{d_{\mathrm{nn}}} |\Omega^{\iota}_{\alpha,\beta,(n)}|^2 \Delta t.
\end{equation}
The matrix $\Omega^{\iota}_{(n)}=(\Omega^{\iota}_{\alpha,\beta,(n)})_{1\leq\alpha,\beta\leq d_s}$ is given by 
\begin{equation}\label{omega}
    \Omega^{\iota}_{(n)}= Q^{\iota}_{(n)}  - \sum_{l=0}^{n-1}  MK_{(l+\frac{1}{2})} C^{\iota}_{(n-l-1)}\Delta t,
\end{equation}
with $n\geq 1$. $\Omega^{\iota}_{(0)}=Q^\iota_{(0)}$ is excluded from Eq.~(\ref{loss-discrete}) because it does not depend on the memory kernel.
 
Similar to the derivation of Eq.~(\ref{ls1}) from Eq.~(\ref{loss-local2}), the optimal condition for the memory kernel is derived from Eq.~(\ref{loss-discrete}), giving 
\begin{equation}\label{cls1}
       \sum_{l=0}^{m_K-1} \underbrace{MK_{(l+\frac{1}{2})}\Delta t}_{U_{l}} 
        \underbrace{\sum_{\iota=0}^{d_{\mathrm{nn}}}
       \sum_{n=0}^{m_K-1} 
       C^{\iota}_{(n-l)}(C^\iota_{(n-k)})^T \Delta t }_{V_{l,k}}
       =
       \underbrace{\sum_{\iota=0}^{d_{\mathrm{nn}}}  \sum_{n=0}^{m_K-1}   
       Q^{\iota}_{(n+1)}(C^\iota_{(n-k)})^T \Delta t}_{W_{k}}.
\end{equation}
To make implementations with standard linear algebra routines evident, in Eq.~(\ref{cls1}) we further define the $d_s\times d_s$ matrices $U_l$,  $V_{l,k}$ and $W_k$. $V_{l,k}$ and $W_k$ are obtained directly from the data, while $U_l$ is to be found with a tensor least square routine. 

For the special case for which $d_s=1$, $U_l$,  $V_{l,k}$ and $W_k$ are just scalars, and we define the vector $U=(U_0,\cdots,U_{m_K-1})$, the matrix $V=(V_{l,k})_{0\leq l\leq m_K-1; 0\leq k\leq m_K-1}$, and the vector $W=(W_0,\cdots, W_{m_K-1})$ . Note that $V$ is positive semi-definite. This is the simplest case
of local kernel approximation, for which $U=V^{-1}W$.

\subsection{Noise generator}\label{app:gar}
Under local kernel approximation, the memory kernel is one-body, but the time evolution of $R$ can not be reduced to a one-body equation of motion. Then, the univariant GAR model should be generalized to the multi-dimensional case.
Let $\bm{R}_{i,(n)}=(R_{i1,(n)},\cdots, R_{id_s,(n)})$ be the $d_s$-dimensional noise acting on site-$i$ at time step $n$. Let $n_A$ be any positive integer.  Let $(\phi_{\alpha,k})_{1\leq \alpha\leq d_s, {1\leq k\leq n_A}}$ be a $d_s\times n_A$ real matrix . 
Assuming translational symmetry, the multi-dimensional GAR model for predicting $\bm{R}_{i,(n)}$ from the history of noise is given by
\begin{equation}\label{1ag}
    \bm{R}_{i\alpha,(n)}  = \sum_{k=1}^{m_A} \phi_{\alpha,k}  R_{i\alpha,(n-k)}  + \sum_{j\in \mathcal{N}(i)} \mu^j_{i\alpha, (n)}   +  \sigma_{i\alpha} w_{i\alpha,(n)}
\end{equation}
and
\begin{equation}
\mu^j_{i\alpha, (n)}= \mu^j_{i\alpha}\big(R_{j\alpha,(n-1)},\cdots,R_{j\alpha,(n-m_A)}\big).    
\end{equation}
Here, $\phi_{\alpha,k}$ are the Yule-Walker linear autoregressive parameters ~\cite{theodoridis2015machine} for $R_{i\alpha,(n)}$ as a time series with respect to $n$. 
$\sigma_{i\alpha}=\sigma_{\alpha}$ is a scalar parameter. In principle, $\sigma_{i\alpha}$ could be a function of the history of $R_{i\alpha,(n)}$, but in all the cases we studied we found that assuming a scalar $\sigma_{i\alpha}$ was sufficient for accurate autoregression, which is the choice that we adopt in the following. 
$w_{i\alpha,(n)}$ is a standard Gaussian white noise, independently sampled for each $i$, $\alpha$ and $n$. $\mathcal{N}(i)$ is the neighborhood of site-$i$. Typically, we can let $\mathcal{N}(i)$ includes site-$i$ and all site-$j$ satisfying $A_{ij}=1$. It is also feasible to include second nearest neighbor beyond $A_{ij}=1$ in $\mathcal{N}(i)$. 
$\mu^{j}_{i\alpha}$ is a one-body function taking as input the history of $R_{j\alpha,(n)}$ as a time series with respect to $n$. 
In ab initio GLE, we use a collection of small feed-forward neural networks to represent $(\mu^{j}_{i\alpha})_{1\leq \alpha\leq d_s}$. The size of the collection can be reduced when system symmetry is taken into consideration. 
For a translationall-invariant lattice, $(\mu^{j}_{i\alpha})$ is decided by the separation between site-$i$ and site-$j$. So the complexity of the GAR model does not grow with the lattice size.

Training of the multi-dimensional GAR model is similar to that of the univariant GAR model. With a given $K^*$, we extract the time series $\bm{R}_{(n)}$ from 
\begin{equation} 
    \bm{R}_{(n)}=M\bm{a}_{(n)} - \bm{\mathcal{F}}_{(n)} - \sum_{s=0}^{n-1}  MK^*_{(s+\frac{1}{2})} \bm{v}_{(n-s-\frac{1}{2})} \Delta t.
\end{equation}
Using $\{ \bm{R}_{(n)} \}$ as data, we first determine the Yule-Walker solution for $\phi_{\alpha,k}$ using a least square routine. Then, we train the parameters in $(\mu^j_{i\alpha})_{1\leq \alpha\leq d_s}$ and $(\sigma_{\alpha})_{1<\alpha<d_s}$ using the maximum likelihood loss function 
\begin{equation}\label{gar-loss-multi}
\begin{split}
    \Gamma &= 
    \sum_{i\alpha} \sum_n \log \sigma_{\alpha}^2  \\
    &+ \frac{
    \big(R_{i\alpha,(n)}
    -\sum_{k=1}^{m_A} \phi_{\alpha,k}R_{i\alpha,(n-k)}
    -\sum_{j\in \mathcal{N}(i)} \mu^j_{i\alpha, (n)}
    \big)^2
    }{
    \sigma_{\alpha}^2
    }
\end{split}
\end{equation}

After training, one can run the GAR model independently, and calculate the noise ACF
$\Upsilon_{(n)}=\langle (\bm{M^{-1}R})_{(0)} (\bm{M^{-1}R})^T_{(n)}\rangle_{\mathrm{GAR}}$. The subscript ``GAR"  indicates that the ensemble average is calculated over trajectories generated by the GAR simulation. 
Then, the ACF of the noise can be compared with $\langle \bm{v}_{(0)}\bm{v}_{(0)}^T \rangle {K^*}^T_{(n)}$ to verify that the 2FDT is satisfied. For that purpose, the equal time velocity correlation matrix $\mathcal{G}=\langle \bm{v}_{(0)}\bm{v}_{(0)}^T \rangle $ is extracted from the data. A small deviation
of $\Upsilon_{(n)}$ from $\mathcal{G} {K^*}^T_{(n)}$ should be expected, reflecting a slight violation of the 2FDT derived from the data. This is an inevitable consequence of the approximation made for the memory kernel, i.e., the local kernel approximation. Thus, the equal-time velocity correlation matrix from the AIGLE simulation will deviate slightly from the corresponding matrix from the MD data.  One may try to reduce this error by modifying the memory kernel to enforce the 2FDT directly. We call $K_*$, a modified memory kernel that minimizes the target function 
\begin{equation}\label{eq:2fdt_loss}
     \mathcal{L}^{\mathrm{FDT}}[K] = L \sum_{n=1}^{m_K} \sum_{\alpha,\beta} \sum_{\iota=0}^{d_{\mathrm{nn}}} \Big(\sum_{\gamma} \mathcal{G}^{\iota}_{\alpha,\gamma}{K}_{\beta,(n)}^{\gamma}-\Upsilon^{\iota}_{\alpha,\beta,(n)} \Big)^2,
 \end{equation}
 where we  follow the conventions of  subsection~\ref{sec:dgle:formulation}, i.e.,  $\mathcal{G}^{\iota}_{\alpha,\gamma} = \frac{1}{L} \sum_i \mathcal{G}_{i\alpha}^{i_{[\iota]}\gamma}$,
 and $\Upsilon^{\iota}_{\alpha,\beta,(n)}=\frac{1}{L} \sum_i\Upsilon_{i\alpha,(n)}^{i_{[\iota]}\beta}$. Thus, 
 $K_*=\mathrm{argmin}_K \mathcal{L}^{\mathrm{FDT}}[K]$, and can be found straightforwardly with a least square scheme. $K_*$ is different from $K^*$, defined in subsection B. In practice, the difference between the two approximate kernels is mostly cosmetic. While $K^*$ is more faithful to the data in terms of the time series regression, $K_*$ is more faithful to the data in terms of the equal-time velocity correlation matrix. In practice, the two approaches produce minor quantitative differences, but are equivalent in terms of physical accuracy.


\section{Details of modeling domain wall motion as a virtual particle}\label{sec-pto}
\subsection{ DFT-based atomistic models}\label{sec-pto-dp}
Our atomistic model for \pto is based on the methodology developed in  Refs.~\cite{PhysRevLett.120.143001, zhang2018end, PhysRevB.102.041121, xie2022ab}. The DP model for representing the potential energy surface is trained on DFT data with the SCAN meta-GGA functional~\cite{sun2015strongly}. 
The dataset is collected through the active learning procedure implemented in the DPGEN code~\cite{zhang2019active, ZHANG2020107206}. The data are labeled by Quantum ESPRESSO ~\cite{giannozzi2009quantum} with norm-conserving pseudo-potentials~\cite{hamann1979norm} including semi-core states. An early version of the current DP model was introduced in Ref.~\cite{xie2022ab}. It was trained with configurations without domain walls and used to study the ferroelectric phase transition in bulk \pto with atomistic simulations. The model predicted thermodynamic and ferroelectric properties of \pto in close agreement with experimental results, and provided unique insight into the microscopic mechanism driving the phase transition. To train the current DP model we follow the same numerical protocol adopted in Ref.~\cite{xie2022ab}, where the full technical details can be found.
\begin{figure}[thb]
    \centering
    \includegraphics[width=0.6\linewidth]{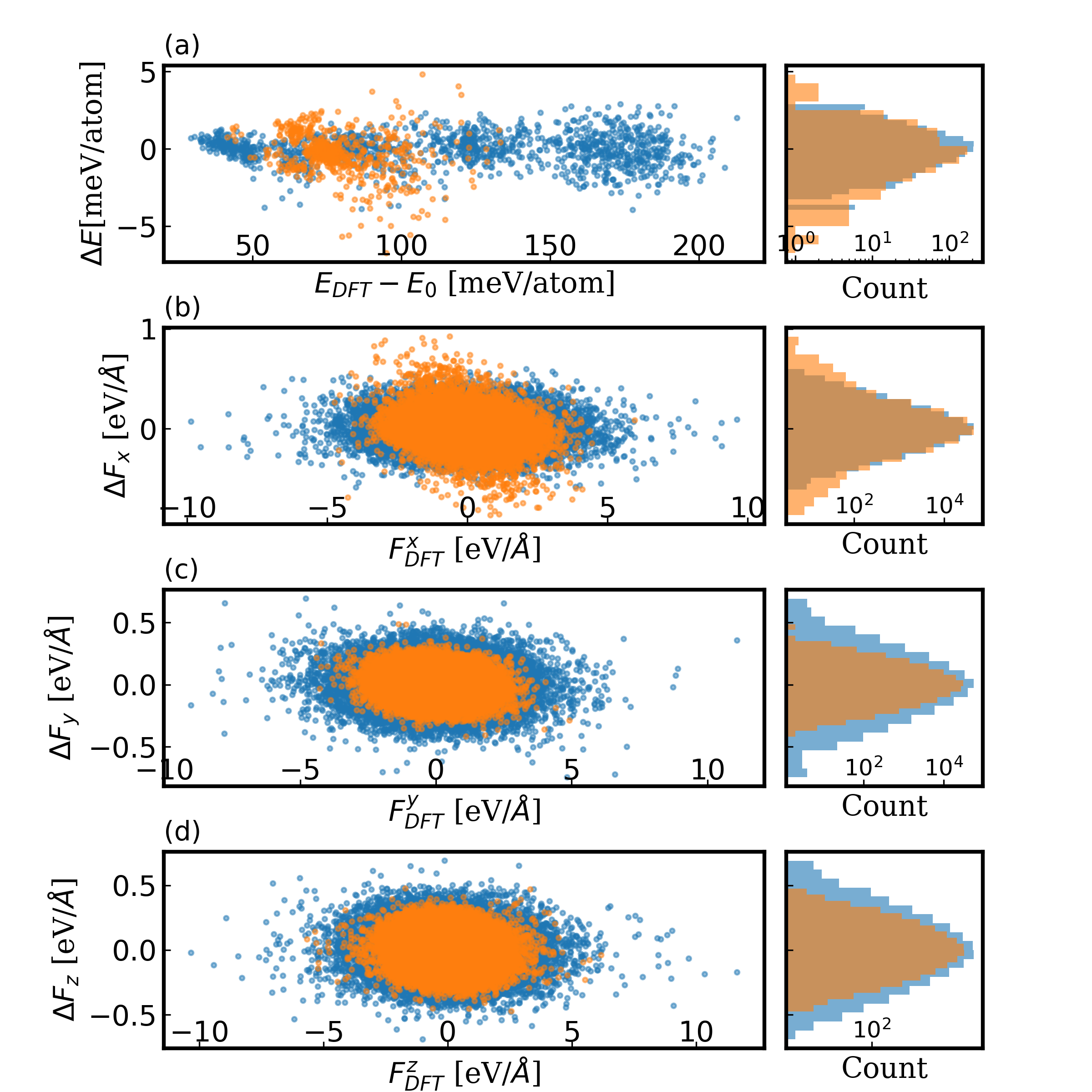}
    \caption{Error distribution of the DP model on the training set. The blue color marks configurations without domain walls, and the orange color marks the rest.  $E_{\text{DFT}},F^{x,y,z}_{\text{DFT}}$ are the energy and force labels of the data. $E_0$ is a constant. $\Delta E, \Delta F_{x,y,z}$ are the difference between the model prediction and the data label.  }
    \label{fig:dperror}
\end{figure}
The final dataset for the current DP model contains configurations without domain walls and with two twin 90${}^{\circ}$ domain walls or two twin 180${}^{\circ}$ domain walls. 
1276 configurations with no domain walls are collected with the active learning procedure of Ref.~\cite{xie2022ab} in the temperature interval  [300K, 1200K] and the pressure interval [0, $10^5$Pa]. 494 configurations with domain walls are collected with the same active learning procedure in the temperature interval $ [100K, 600K]$ and the pressure interval [0, $10^5$Pa]. The largest supercell with domain wall configurations contains $54$ elementary cells, corresponding to 270 atoms. We use a spatial cutoff radius of $8\mathrm{\AA}$  for the DP model.  This is larger than in typical DP models but is necessary to capture long-range interactions in the presence of two domain walls with periodic boundary conditions.  
 
Fig.~\ref{fig:dperror} plots the nearly Gaussian error distribution of the DP model against the training set. The root-mean-square error (RMSE) of the energy is 1meV/atom for configurations without domain walls and 1.4meV/atom for the remaining ones. The RMSE of the Cartesian components of the forces is around 0.1eV/$\mathrm{\AA}$  in all cases. The force error is slightly larger for the direction orthogonal to the domain walls, which is the x-axis for our data with 180${}^{\circ}$ domain walls. The DP model is also validated on a small independent testing set, showing similar accuracy to its performance on the training set. We conclude that the DP model reproduces statistically the adiabatic potential energy surface of SCAN-DFT. 

Next, we calculate the Berry-phase polarization change, relative to the centrosymmetric structure for all configurations without domain walls. The calculation is done with the Wannier90 code ~\cite{pizzi2020wannier90}. The total polarization is the sum of electronic (Berry phase) contributions and ionic contributions~\cite{resta2007theory}. The dipole $D$ of the simulation cell (cell dipole) is obtained by multiplying the total polarization by the volume of the cell. The  Born charge (BC) associated with atom-$i$ is the tensor $Z_{i,\alpha\beta}=\frac{\partial D_\alpha}{\partial x_{i,\beta}}$ defined as the change of $D$ in direction $\alpha$ caused by the displacement of atom-$i$ away from its equilibrium position in direction $\beta$. We adopt a linear approximation for the dependence of $D_\alpha$ on the atomic coordinates, given by  $D_\alpha=\sum_i \sum_{\beta\in \{x,y,z\}}Z_{i,\alpha\beta} x_{i,\beta}$, which, in the manuscript, is referred to as the BC model.  
Due to space-group symmetry or negligible contribution, some degrees of freedom of the BC model can be eliminated. For \pto, the BCs of Pb and Ti atoms are conventionally approximated~\cite{PhysRevB.55.6161} by diagonal matrices $Z_{\mathrm{Pb},\alpha\beta}=Z_{\mathrm{Pb}}\delta_{\alpha\beta}$ and $Z_{\mathrm{Ti},\alpha\beta}=Z_{\mathrm{Ti}}\delta_{\alpha\beta}$, respectively. The BC of the O atom is also given a diagonal approximation, with values $Z_{\mathrm{O1}}$ for the Ti-O bond direction and $Z_{\mathrm{O2}}$ for the other two orthogonal directions. 
In the present study, the BC model is fitted to the cell dipole data with mean squared error loss, leading to $Z_{\mathrm{Pb}}=3.7140e$, $Z_{\mathrm{Ti}}=5.4897e$, $Z_{\mathrm{O1}}=-3.3551e$ and $Z_{\mathrm{O2}}=-2.9234e$. The RMSE of the cell dipole predicted by the resultant BC model is 2e$\mathrm{\AA}$  for a $3\times3\times3$ supercell. This corresponds to roughly a $2\mu \mathrm{C/cm}{}^2$ RMSE in polarization, much smaller than the  $72\mu \mathrm{C/cm}{}^2$ total polarization of \pto extracted from our room temperature simulations. The BC model constructed in this way is statistically more accurate at finite temperature than models using conventional Born effective charges calculated by perturbing the ground state equilibrium structure~\cite{zhong1995first}. 
Using the BC model so defined, the local electric dipole $p_j$ associated with each Ti-centered elementary cell-$j$ of \pto is given by a weighted sum of the position of the atoms in the cell-$j$ ~\cite{meyer2002ab,behera2011structure}. The local dipole defined via the BC model is an approximation of the local dipole introduced in Ref.~\cite{xie2022ab} via maximally localized Wannier functions. The latter includes full (nonlinear) environmental dependence of the local dipoles, which is computationally significantly more expensive than the simple BC model. So in the current study, we use the BC model, to accelerate the simulation of domain wall dynamics.  We also remark that the linear approximation is sufficiently accurate for MD simulations at room temperature, which is sufficiently lower than the ferroelectric phase transition temperature of \pto.  
 \begin{figure}[thb]
    \centering
    \includegraphics[width=0.5\linewidth]{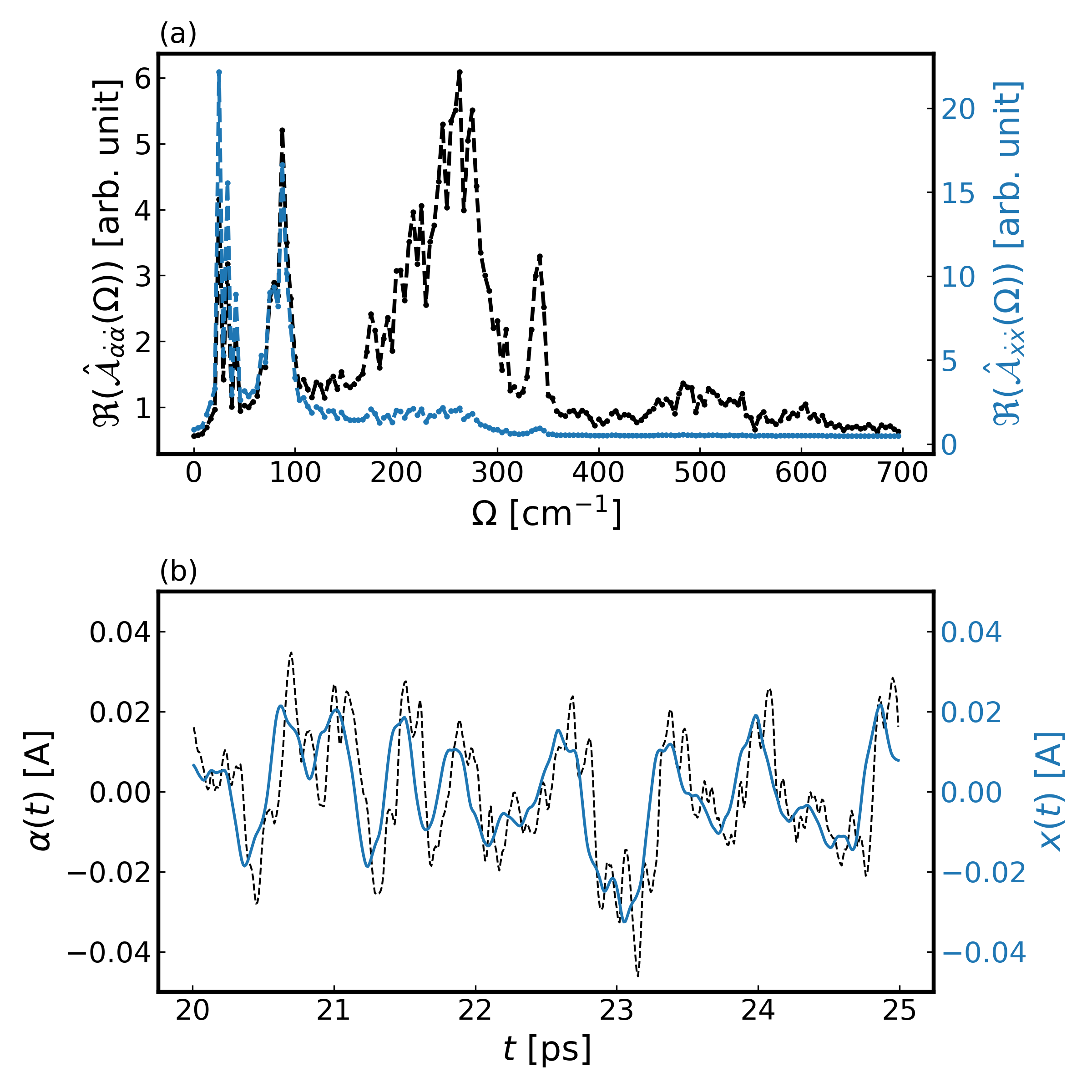}
    \caption{(a) The real parts of $\hat{\mathcal{A}}_{\dot{\alpha}\dot{\alpha}}(\Omega)$ and $\hat{\mathcal{A}}_{\dot{x}\dot{x}}(\Omega)$. (b) A segment of a MD trajectory of $\alpha$ (black) under zero external field. The corresponding trajectory of $x$ (blue) in the same time interval is plotted as the blue line for comparison. }
    \label{fig:covv}
\end{figure}

\subsection{MD simulations}
With the BC model,  electric field-driven (non-equilibrium) MD can be simulated from the evolution under a  Hamiltonian $H$ obtained by adding the interaction of the dipoles with the external field $E$ to the Hamiltonian $H_0$ in the absence of the field, i.e.,  $H = H_0 - \sum_j \bm{E}\cdot \bm{p_j}$.  
We use LAMMPS~\cite{LAMMPS} and PLUMED~\cite{tribello2014plumed} for these simulations. 

In our MD simulations, we use a $20\times 40\times 20$ supercell of \pto. 
The cell tensor is diagonal and kept fixed along the $x$, $y$ direction (lattice constant $a=b=3.91\mathrm{\AA}$) to mimic the experimental lattice constants in the plane, imposed by the substrate on epitaxially grown thin films. So the area of the $20\times 40$ $xy$ plane of the supercell is fixed as $122.305\mathrm{nm}^2$. This arrangement mimics \pto thin films deposited on  SrRu$\text{O}_3$/SrTi$\text{O}_3$ substrates~\cite{morita2004epitaxial, KIM201447, nishino2020evolution, Dahl2009}.
We approximate the environmental noise by applying a Langevin thermostat to the thin atomic layers highlighted in pink in Fig.1 of the main text, which contain atoms far away from the initial locations of the domain walls.
A barostat is applied in the $z$ direction only. The target pressure $P_z=2.8\times 10^4$ bar is determined by roughly matching the calculated average lattice constant $c$ to the experimental value under room temperature and atmospheric pressure. A larger than atmospheric pressure reduces the super-tetragonality error of the adopted DFT functional approximation (see Ref.~\cite{xie2022ab,zhong1995first} for more details).  This setup approximates a thin \pto film between conducting electrodes. Notice, however, that the adoption of periodic boundary conditions rules out a precise connection between our model and experimental ultra-thin \pto samples, where depolarization effects play an important role ~\cite{dawber2003depolarization}. Another difference between our model and experiment is that the finite size of the supercell along the $x/z$ directions reduces the nucleation-driven growth of the domain~\cite{shin2007nucleation}.

\subsection{Temporal coarse-graining}
 The velocity autocorrelation function (ACF) of the CV $\alpha$ is $\mathcal{A}_{\dot{\alpha}\dot{\alpha}}(\tau)=\langle \dot{\alpha}(\tau)\dot{\alpha}(0)\rangle$. We use $\hat{\mathcal{A}}_{\dot{\alpha}\dot{\alpha}}(\Omega)$ to denote its Fourier transform. Similarly, we use $\hat{\mathcal{A}}_{\dot{x}\dot{x}}(\Omega)$ to denote the Fourier transform of the velocity ACF of the temporally coarse-grained CV $x$. The real parts of $\hat{\mathcal{A}}_{\dot{\alpha}\dot{\alpha}}(\Omega)$ and $\hat{\mathcal{A}}_{\dot{x}\dot{x}}(\Omega)$ are plotted in Fig.~\ref{fig:covv}(a). Through the temporal coarse-graining from $\alpha$ to $x$, we see the damped modes of $\alpha$ peaked near $280\mathrm{cm^{-1}}$, $340\mathrm{cm^{-1}}$, $500\mathrm{cm^{-1}}$ and $600\mathrm{cm^{-1}}$ are suppressed. Among all, the mode near $280\mathrm{cm^{-1}}$ is not completely eliminated but becomes a continuous spectrum between $100\mathrm{cm^{-1}}$ and $400\mathrm{cm^{-1}}$. The magnitude of the continuous spectrum is, however, negligible compared to the slow modes between $0\mathrm{cm^{-1}}$ and $100\mathrm{cm^{-1}}$. So the integration time step of AIGLE with respect to $x$ is mainly limited by the mode peaked near $90\mathrm{cm^{-1}}$.

 To have a more straightforward view of temporal coarse-graining, we plot  $\alpha(t)$ and $x(t)$ for the same segment of MD trajectory (no domain motion event) simulated under zero external electric field, in Fig.~\ref{fig:covv}(b). It is clear the fast and small oscillation of $\alpha$ is reduced in $x$, while the slow and strong oscillation roughly between $0.02A$ and $-0.02A$ is preserved. 

\subsection{ Training of AIGLE}
In the first two training steps of the AIGLE model, the dataset consists of several $400$ps long trajectories of $x$ under $E=2$mV/{\AA}. The underlying Hamiltonian dynamics is at metastable equilibrium --- the twin domain walls are trapped by the periodic potential around a local minimum within the simulated time scale. Under such circumstance, the potential energy surface of the CV is not fully explored, but the fast atomistic degrees of freedom are sufficiently ergodic for determining $K$ and the GAR model. The GAR model contains a feed-forward neural network with two hidden layers (size=10). The $\sigma_{(n)}$ is parameterized by one scalar variable. We use the Adam optimizer with an initial learning rate of 0.01 and an exponential decay rate of 0.9  every 500 steps. We let  $n^{\mathrm{GD}}=10$ and $\epsilon=0.01$. Convergence is reached within 5000 iterations. The entire dataset is used for every iteration.  After training, the colored noise $R$ and the white noise $w$ are both distributed normally on the dataset with negligible off-centering. Non-stationarity is not detected numerically for the GAR model.  

In the out-of-equilibrium regime, we fix the memory kernel and the GAR model. Then we retrain the force field $\mathcal{F}(x) = -\partial_x U(x) + pEx$ with MD trajectories  (about $  1.2$ns long in total) simulated independently under different $E\in [2.0,2.4]$mV/A. In this dataset, MD can be out-of-equilibrium by creeping down metastable states. Although the entire landscape is explored, the creeping events are too sparse in the dataset for fitting the barrier height $U_b$ accurately. So we predetermine it using metadynamics, a method for computing free energy differences ~\cite{barducci2008well},  under $E=0$. The resulting free energy barrier $\Delta E$ in energy units is converted to $U_b=\frac{\Delta E a^2}{m}$, where $a=3.91\mathrm{\AA}$  is the lattice constant.  All the other parameters in $\mathcal{F}(x)$ are then trained directly on the MD dataset. We use the Adam optimizer with the same setting. Convergence is reached within 5000 iterations. The entire dataset is used at each iteration.

\begin{figure*}[tb]
    \centering
    \includegraphics[width=0.6\linewidth]{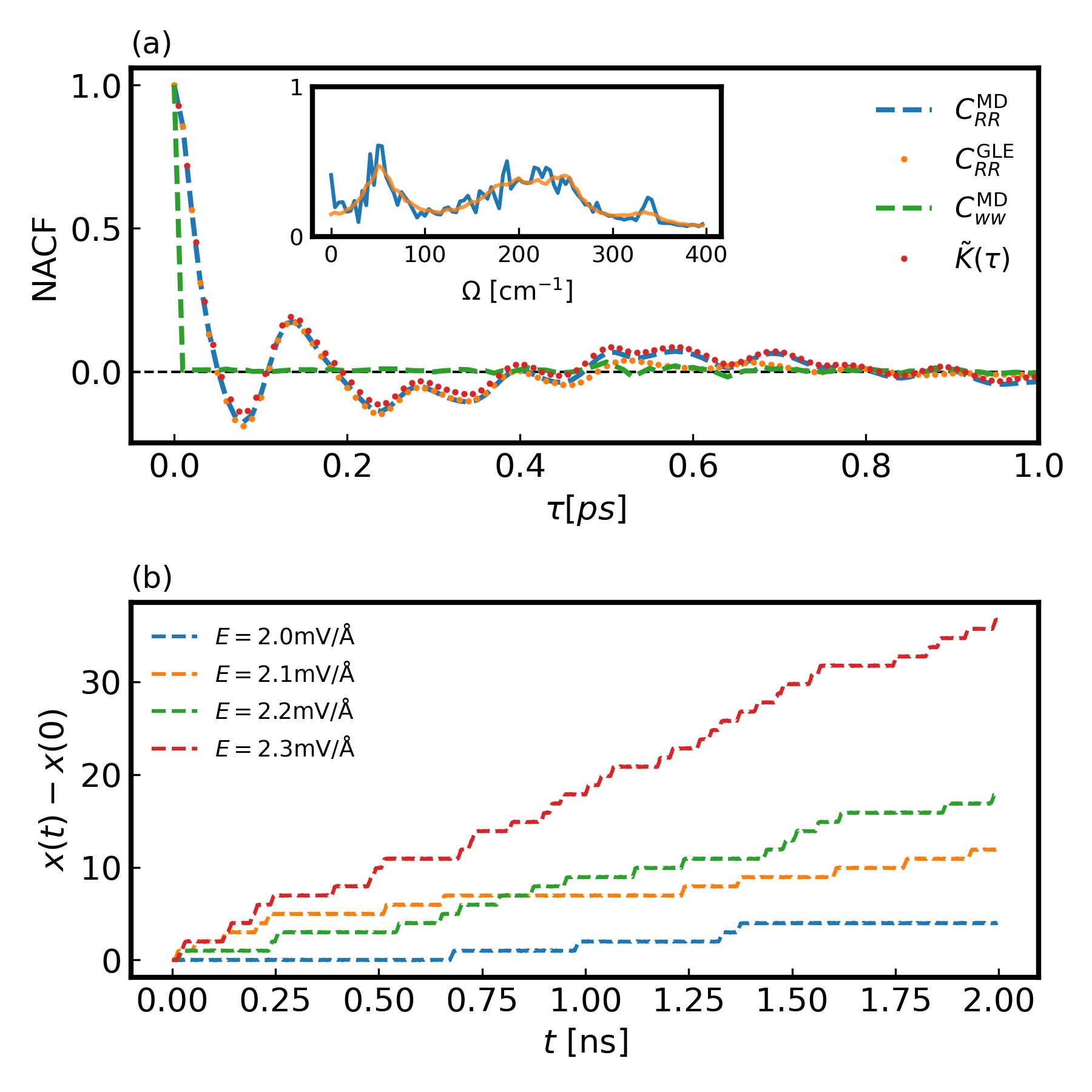}
    \caption{(a) The NACF of the noise and the rescaled memory kernel $\tilde{K}(\tau)$ for $\tau<1$ps. For  $\tau>1$ps correlations exhibit a weakly oscillatory pattern similar to that shown here in the vicinity of $\tau=1$ps. The inset shows $\Re(\hat{C}^{\text{MD}}_{RR}(\Omega))$ (blue) and $\Re(\hat{C}^{\text{GLE}}_{RR}(\Omega))$ (orange) in arbitrary units. $\Re(\hat{C}^{\text{MD}}_{RR}(\Omega))$ displays unphysical oscillations originating from statistical errors due to limited MD data. The first peak in $\Re(\hat{C}^{\text{GLE}}_{RR}(\Omega))$ is located near $50\mathrm{cm^{-1}}$.  (b) Time evolution of the CV $x$ under different driving fields $E$. The waiting time distribution for the domain motion has a long tail.}
    \label{fig:nacf}
\end{figure*}
The productive AIGLE model is trained through the steps described above. Next, we examine its agreement with the statistics of MD data. We first extract the reference noise from the MD data with our AIGLE model. Then we use the GAR model to generate the simulated noise time series for comparison. 
In Fig.~\ref{fig:nacf}(a), we report the normalized autocorrelation function (NACF) of the reference noise extracted from the MD data at metastable equilibrium, defined as $C^{\text{MD}}_{RR}(\tau)=\langle R(t_0+\tau)R(t_0)\rangle/\langle R(t_0)^2\rangle$, and the NACF of the corresponding reference residual noise $C^{\text{MD}}_{ww}(\tau)=\langle w(t_0+\tau)w(t_0)\rangle/\langle w(t_0)^2\rangle$, which is also extracted from the MD data. It is apparent that the cutoff $m_A$ is large enough since $C^{\text{MD}}_{RR}(\tau>m_A\Delta t)\ll 1$. Moreover, 
 $C^{\text{MD}}_{ww}(\tau)$ is delta-like, except for a very small residual correlation for $\tau > m_A\Delta t$.  Additionally, both $R$ and $w$ display zero-centered normal distributions on the dataset, indicating that the GAR model successfully reduces the correlated noise in MD trajectories into Gaussian white noise. The NACF of the simulated noise (generated by the GAR model), denoted by $C^{\text{GLE}}_{RR}(\tau)$, is also plotted in Fig.~\ref{fig:nacf}(a). $C^{\text{GLE}}_{RR}(\tau)$ captures accurately the dominant oscillatory feature of $C^{\text{MD}}_{RR}(\tau)$,
 but for a small discrepancy for $\tau > m_A\Delta t$, which is caused by the residual correlation not eliminated by GAR. To display this effect in another way, we compare the Fourier transform of $C^{\text{MD}}_{RR}(\tau)$ and of $C^{\text{GLE}}_{RR}(\tau)$, denoted by 
$\hat{C}^{\text{MD}}_{RR}(\Omega)$ and by $\hat{C}^{\text{GLE}}_{RR}(\Omega)$, respectively. The corresponding real parts, $\Re(\hat{C}^{\text{MD}}_{RR}(\Omega))$ and $\Re(\hat{C}^{\text{GLE}}_{RR}(\Omega))$, shown in the inset of Fig.~\ref{fig:nacf}(a), display overdamped modes within $[0,400] \mathrm{cm}^{-1}$. As expected, the noise captures mainly broad overdamped modes, due to the interactions of the CV with a bath of fast variables. 
The main difference between $\Re(\hat{C}^{\text{MD}}_{RR}(\Omega))$ and $\Re(\hat{C}^{\text{GLE}}_{RR}(\Omega))$ occurs for frequencies close to zero, associated to long-time correlation of the noise that can not be captured accurately by a GAR model with finite memory. Given that long-time correlations appear to be weak and fluctuating in $C^{\text{MD}}_{RR}(\tau)$, the errors due to neglecting them in the GAR model should have a negligible overall impact. 
Finally, in Fig.~\ref{fig:nacf}(a), the optimized memory kernel $K$ is rescaled as in $\tilde{K}(\tau)= K(\tau)\frac{C^{\mathrm{MD}}_{RR}(\Delta t/2)}{K( \Delta t/2)}$ to facilitate direct comparison with $C^{\text{MD}}_{RR}(\tau)$. The excellent agreement observed in the figure indicates that the 2FDT is well-satisfied by AIGLE at metastable equilibrium.   

The productive AIGLE model is able to simualte efficiently the dynamics of the CV under finite electric field $E$. 
Segments of CV trajectories simualted by AIGLE are plotted in Fig.~\ref{fig:nacf}(b), where one can see that the time scale of domain displacement events increases from picoseconds to nanoseconds with a modest decrement of $E$. 

\subsection{Ferroelectric switching}
 \begin{figure}[tb]
    \centering
    \includegraphics[width=0.4\linewidth]{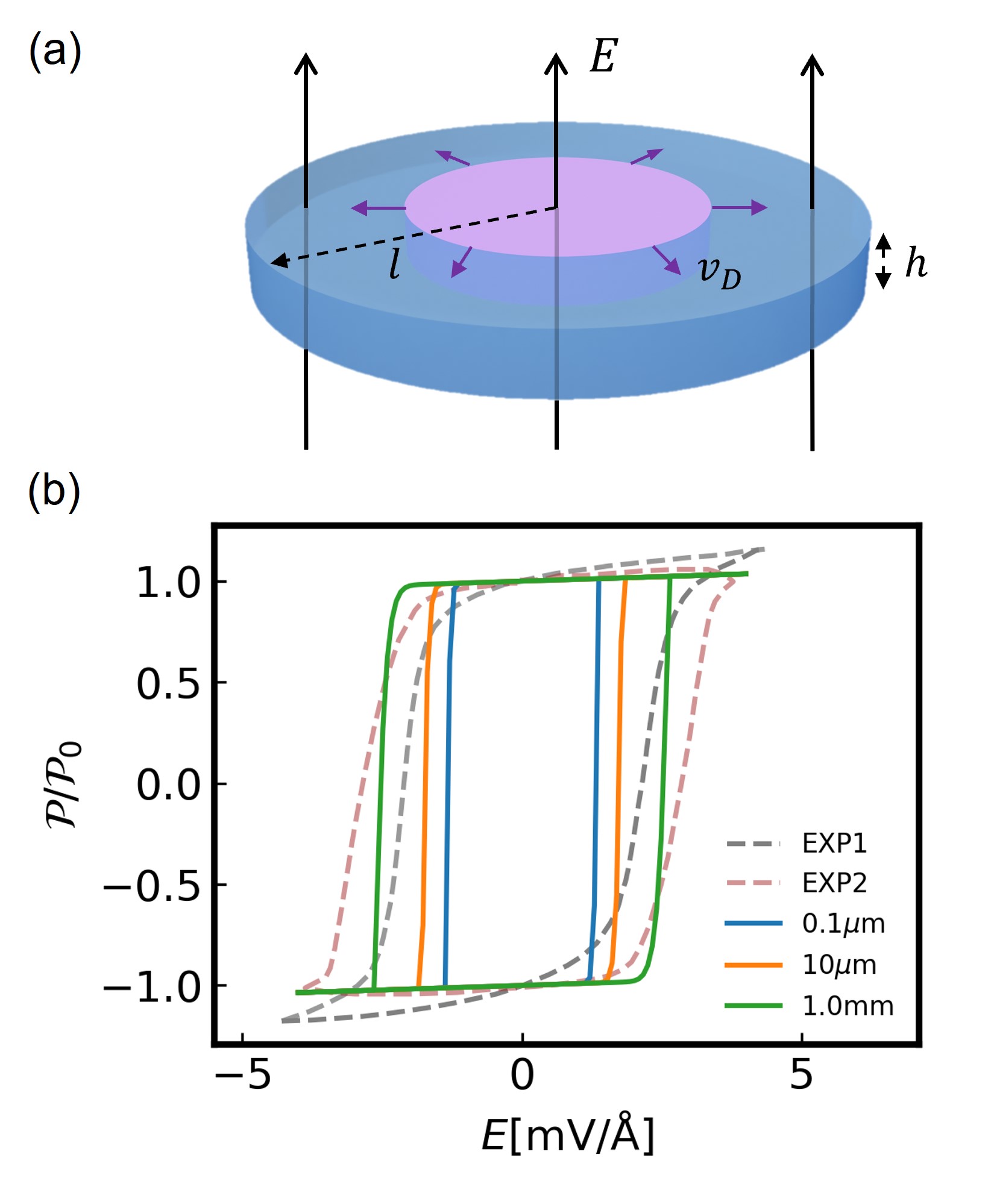}
    \caption{  (a) Schematic representation of the sideways-growth stage of ferroelectric switching. In our model, the effects of the finite film thickness (h) are ignored. The radius of the thin film is $l$. The purple domain has upward polarization, parallel to the external field $E$. The blue domain has downward polarization. The speed of growth $v_D$ can be approximated by the speed of a flat domain wall under the same external field, when the upward domain is sufficiently large.   (b) The hysteresis loop $\mathcal{P}$, scaled by $1/\mathcal{P}_0$, for $t_p=1/f=100\mu$s and different $l$. $\mathcal{P}_0$ is the remnant polarization at $E=0$. The ``EXP1'' experimental data ~\cite{KIM201447} were obtained for the same $t_p$ and unknown $l$. The ``EXP2'' data ~\cite{Morita2004} were obtained for $l\approx 144 \mu$m and unknown $t_p$. The remnant polarization in the two experiments is $94\mu \mathrm{C/cm}{}^2$~\cite{KIM201447} and $96\mu \mathrm{C/cm}{}^2$~\cite{Morita2004}, respectively. Our atomistic model gives $\mathcal{P}_0=72\mu \mathrm{C/cm}{}^2$.}
    \label{fig:switch}
\end{figure}
The switching process of a ferroelectric thin film typically involves three stages. The first stage is the nucleation of opposite domains at particular nucleation sites. The second stage is the forward growth of the nucleus across the thin film. The last stage is the sideways growth (widening) of the domain. For perovskite oxides, the first two stages are usually much faster than the last one~\cite{dawber2005}. Hence, as a rough approximation, one can consider just the sideways growth of a cylindrical domain for computing the hysteresis loop of a thin film of \pto. 
Moreover, when the radius of the switched area is much larger than several nanometers, the curvature dependence of the domain wall velocity should be negligible. 

 Here, we consider a thin \pto film of radius $l\in[1,1000]\mathrm{\mu m}$, initially polarized downward, as shown in Fig.~\ref{fig:switch}(a).  An upward square electric field pulse of magnitude $E$ and duration $t_p$ is applied to the film, causing the growth of a central cylindrical domain with upward polarization separated by an interface from the surrounding medium with downward polarization.
Alternatively, one could also consider the growing domain to be cuboidal instead of cylindrical. In our atomistic simulations, we find a cuboidal domain, with round corners, is structurally more stable than its cylindrical counterpart when growing (not true when it is shrinking). 
In this case, approximating $v_D$ with the one computed from an infinite planar domain wall will only be more appropriate.
 
Because $l$ is large enough, the domain wall velocity can be well approximated by the $v_D$ computed for infinitely large flat domain walls. So after the application of the square pulse, the radius of the upward domain is $r_{\mathrm{\uparrow}}=\min(v_D(E)t_p, l)$. The cross-sectional area is then $A_\uparrow=\pi r_{\mathrm{\uparrow}}^2$ for the upward domain, and $A_\downarrow=\pi l^2 - \pi r_{\mathrm{\uparrow}}^2$ for the downward domain. 
The lower branch of the hysteresis loop $\mathcal{P}(E)$ can be approximated by the average polarization of the two domains: 
\begin{equation}\label{eq:hys}
\mathcal{P}(E)=\frac{A_\uparrow}{\pi l^2}(\mathcal{P}_0+\chi E) +  \frac{A_\downarrow}{\pi l^2} (- \mathcal{P}_0+\chi E),
\end{equation}
where the remnant polarization  $\mathcal{P}_0=72\mu \mathrm{C/cm}{}^2$ and the susceptibility $\chi=0.643 \mathrm{nC/mV}$ are extracted from the MD simulations. The upper branch of the hysteresis loop is computed in a similar way. 

 With this toy model of ferroelectric switching, the hysteresis loop for three typical domain radii $l$ are reported in Fig.~\ref{fig:switch}(b), where we also display the hysteresis loops observed in two experiments~\cite{KIM201447, Morita2004}.
 The vertical span of the loops is determined by the remnant polarization $\mathcal{P}_0$, which is equal to 72$\mu \mathrm{C/cm}^2$ in our atomistic model and is equal to 94$\mu \mathrm{C/cm}^2$ and 96$\mu \mathrm{C/cm}^2$, respectively, in the two experiments. The loop width depends on the radius $l$ of the film, on the velocity $v_D(E)$ of the interface, and on the coercive field associated to the polarization switch. Quite remarkably, simulations without any empirical input are able to reproduce closely the widths of the loops observed in experiments. The main difference between theory and experiment is that the polarization switch is much sharper in the former than in the latter. Interestingly, had we used Merz's law instead of AIGLE to compute $v_D(E)$, the theoretical loops would have been even sharper, albeit marginally so. The effect is small because near the coercive field, the difference between AIGLE and Merz's law is minor. The discrepancy between theory and experiment in the switching rate of the polarization with $E$ suggests that effects beyond our simple model play a role. 
 Realistic models should describe polarization pinning by point defects ~\cite{he2003first}, the finite thickness of the experimental samples~\cite{nagarajan1999thickness, dawber2003depolarization}, edge effects, the curvature dependence of the domain wall velocity, the morphology of the samples where several polarization domains may be present and eventually coalesce, etc. 

\section{Details of modeling coarse-grained lattice dynamics}\label{sec-pto}
\subsection{ DFT-based atomistic models and MD simulations}\label{sec-pto-dd}
For this part of the study, we used the same DP model as the one introduced in  Sec.~\ref{sec-pto-dp}. But for computing the local dipole moments as CVs, we use the Deep Dipole model instead of the effective Born charge model. The latter is understood as a linear approximation to the former. The Deep Dipole model used here is exactly the same as the one introduced and analyzed in detail, in Ref.~\cite{xie2022ab} via maximally localized Wannier functions.  

We carry out two types of MD simulations. The first is $E=0$ NVT-MD simulation of a $8\textrm{nm}\times 8\textrm{nm}\times5\textrm{nm}$ ($20\times 20\times 12$ elementary cells) supercell within a single ferroelectric domain. We collect $30$ps-long trajectories of all local dipole moments as training data. The second type of MD simulations is electric field-driven NVT-MD simulation of a $20\textrm{nm}\times 20\textrm{nm}\times5\textrm{nm}$ ($50\times 50\times 12$ elementary cells) supercell initialized with two opposite ferroelectric domains. 
For $E=0.5$mV/A and $E=1$mV/A, we collect respectively $20$ps-long trajectories of all local dipole moments as training data. In addition, we collect $200$ps-long trajectories of local dipole moments as validation data, used in Fig.3(d,e) of the main text for direct comparison with AIGLE.

 \subsection{ Training of AIGLE}
We first determine the mass of the local dipole moment from the trajectory data $\{\bm{x}_{(n)}\}$  with the equipartition theorem. We train the parameters in the free energy $G$ through a simple force matching scheme: minimizing the mean-squared-error loss $\mathcal{L}^{\mathrm{MSE}}=\sum_n \|M\bm{a}_{(n)}  + \nabla_{\bm{x}} G(\bm{x}_{(n)}) - pM\bm{E}_{(n)}\|^2  $ over the trajectory data collected under zero external field ($E=0$). Then, we calculate the memory kernel $K^*$ through Eq.~(\ref{cls1}), followed by the calculation of the Yule-Walker coefficients in Eq.~(\ref{1ag}). Note that here we do not use any iterative scheme since the free energy is predetermined. Next, with the free energy model and the memory kernel, we extract the noise from the data through $\bm{R}_{(n)} = M\bm{a}_{(n)}-\bm{\mathcal{F}}_{(n)} - \sum_{l=0}^{n-1}  MK_{(l+\frac{1}{2})} \bm{v}_{(n-l-\frac{1}{2})} \Delta t$. We train the GAR model, with $\bm{R}_{(n)}$ as the time series data and the Yule-Walker coefficients frozen, through the maximum likelihood loss Eq.~(\ref{gar-loss-multi}). After the training of the GAR model, we reinforce the 2FDT by modifying the memory kernel from $K^*$ to $K_*$, as given by Eq.~(\ref{eq:2fdt_loss}). In the end, we retrain the parameters of the free energy and the response coefficient $p$ by minimizing the loss 
\begin{equation}
    \mathcal{L}^{\mathrm{MSE}'} =\sum_n \frac{ \sum_i \|\Delta F_{i,(n)} \|^2 \Theta(|x_{i,(n)}|-x_0)}{ \sum_i \Theta(|x_{i,(n)}|-x_0)}   + \frac{ \sum_i \|\Delta F_{i,(n)} \|^2 \Theta(x_0-|x_{i,(n)}|)}{ \sum_i \Theta(x_0-|x_{i,(n)}|)} ,    
\end{equation}
where $ \Delta {\bm F}_{(n)}=M\bm{a}_{(n)}  + \nabla_{\bm{x}} G(\bm{x}_{(n)}) - pM\bm{E}_{(n)}$, over the trajectory data collected under $E=0$, $E=0.5$mV/A and $E=1$mV/A. $\Theta(|x_{i,(n)}|-x_0)$ is the Heaviside step function that vanishes if the magnitude of the local dipole moment $x_{i,(n)}$ is smaller than a threshold value $x_0=2\mathrm{e\AA}$, which distinguishes a local dipole at the interface of opposite domains from the local dipoles in the bulk. The goal is to improve the accuracy of the force field at the interface since these configurations are not included in the first force-matching step.
After all these steps, we obtain the AIGLE model yielding the results displayed in the main text.

\bibliography{pnas-sample}